\documentclass[
superscriptaddress,
reprint,
 amsmath,amssymb,
aps,
pra,
floatfix]{revtex4-2}

\usepackage{bbold}
\usepackage{graphicx}
\usepackage{dcolumn}
\usepackage{bm}
\usepackage{setspace}
\usepackage{mathrsfs}
\usepackage{hyperref}

\usepackage{graphicx}
\usepackage{dcolumn}
\usepackage{bm}
\usepackage[normalem]{ulem}

\usepackage{amsmath}
\usepackage[caption=false]{subfig}

\usepackage{braket}

\usepackage{pgfplots}
\pgfplotsset{width=8.6cm,compat=1.5}

\usepackage{textcomp}

\usepackage{xcolor}

\usepackage{hyperref}
\hypersetup{
    colorlinks=true,
    linkcolor=blue,
    citecolor=blue,      
    urlcolor=blue,
}

\usepackage{csquotes}

\usepackage{bbold}

\usepackage{mathrsfs}

\usepackage[rightcaption]{sidecap}

\usepackage{floatrow}
\graphicspath{{Pictures/}}

\newcommand{\figref}[1]{Fig.~\ref{#1}}
\newcommand{\bvec}[1]{\boldsymbol{#1}}
\newcommand{\eqnref}[1]{Eq.~\eqref{#1}}

\begin{document}

\preprint{APS/123-QED}

\title{Achieving Sub-Zeptonewton Force Sensitivity and Spin-Motion Entanglement in Levitated Diamond via Pulsed Backaction Evasion }

\author{Gayathrini Premawardhana}
\email{gtp6626@umd.edu}
\affiliation{Joint Center for Quantum Information and Computer Science, University of Maryland-NIST, College Park, Maryland 20742, USA}
\author{Jonathan Beaumariage}
\affiliation{Department of Physics and Astronomy, University of Pittsburgh, Pittsburgh, Pennsylvania 15260, USA}
\affiliation{Pittsburgh Quantum Institute, University of Pittsburgh, Pittsburgh, Pennsylvania 15260, USA}
\author{M. V. Gurudev Dutt}
\affiliation{Department of Physics and Astronomy, University of Pittsburgh, Pittsburgh, Pennsylvania 15260, USA}
\affiliation{Pittsburgh Quantum Institute, University of Pittsburgh, Pittsburgh, Pennsylvania 15260, USA}
\author{David Pekker}
\affiliation{Department of Physics and Astronomy, University of Pittsburgh, Pittsburgh, Pennsylvania 15260, USA}
\affiliation{Pittsburgh Quantum Institute, University of Pittsburgh, Pittsburgh, Pennsylvania 15260, USA}
\author{Thomas Purdy}
\affiliation{Department of Physics and Astronomy, University of Pittsburgh, Pittsburgh, Pennsylvania 15260, USA}
\affiliation{Pittsburgh Quantum Institute, University of Pittsburgh, Pittsburgh, Pennsylvania 15260, USA}
\author{Jacob M. Taylor}
\email{jmtaylor@umd.edu}
\affiliation{Joint Center for Quantum Information and Computer Science, University of Maryland-NIST, College Park, Maryland 20742, USA}
\affiliation{Joint Quantum Institute, University of Maryland-NIST, College Park, Maryland 20742, USA}

\date{\today}

\begin{abstract}

We propose a system to achieve sub-zeptonewton force sensing and robust spin-mechanical entanglement in a levitated diamond system. By coupling a nitrogen-vacancy (NV) center spin to the motion of its host diamond within a magnetic trap, we develop a platform designed to surpass the standard quantum limit. We develop and compare three distinct pulse sequences—Ramsey, Hahn echo, and Carr-Purcell-Meiboom-Gill (CPMG)—to create increasing amounts of backaction evasion while mitigating the effects of shot noise and thermal decoherence. Our results show that the CPMG sequences yield the most significant performance gains, reaching a force sensitivity of better than $10^{-23} \text{ N}/\sqrt{\text{Hz}}$ for broadband sensing around $10^4 \text{ Hz}$. Furthermore, we derive an entanglement witness protocol that accounts for pulsed dynamical decoupling, proving that spin-motion entanglement remains detectable even when occurring much faster than the mechanical period. These findings provide a more practical path for using levitated nanodiamonds both as high-precision sensors and as non-classical mechanical systems for fundamental tests of quantum mechanics.
\end{abstract}

\maketitle

\section{Introduction}

Levitated systems provide a versatile platform for creating low noise mechanical devices with applications ranging from sensing to medicine to fundamental physics \cite{NeukirchEtAl2,NeukirchEtAl1,JinEtAl,VoisinEtAl,HoangEtAl,PerdriatEtAl,RiviereEtAl,gonzalez2021,BorettiNMR}. Of particular interest are levitated systems that may be put into non-classical states, which have promise for testing key ideas in theories at the intersection of quantum mechanics and gravitation \cite{YinEtAl,ScalaEtAl,WoodEtAl,ZhouEtAl,PedernalesEtAl,WanEtAl1,WanEtAl2}. Towards that end, levitated diamonds with one or a few nitrogen-vacancy (NV) centers have been suggested as a system uniquely suited to the task. This optically-accessible spin system has demonstrated a unique combination of magnetic sensitivity and spatial resolution while having good coherence properties even in small (sub-micrometer sized) crystals\cite{MazeEtAl, NusranEtAl,BurglerEtAl,JiangEtAl,KatsumiEtAl,VetterEtAl,ZhangEtAl,DeLeonNV2}.

Here we consider how to take advantage of this platform due to its potential for further decoupling from environmental noise, especially when operated in vacuum \cite{JinEtAl,peng2018thesis}, while maintaining the ability to prepare non-classical states by using the spin of the NV to create a spin-dependent force. Initial demonstrations of diamond levitation in rough vacuum were achieved by Neukirch et al. \cite{NeukirchEtAl2}, who notably encoded the diamond's mechanical motion in the spin fluorescence. Early explorations of optically detected magnetic resonance (ODMR) in high-vacuum environments were conducted in Ref.~\cite{peng2018thesis}, and Jin et al. \cite{JinEtAl} recently demonstrated expanded capabilities, reporting dynamic spin control and Berry phase measurements in high vacuum. These milestones underscore a broader effort within the community to achieve stable spin-mechanical coupling in environments where gas damping is minimized.

Our focus is leveraging the strong coupling achievable for small crystals under the conditions of magnetic levitation, which involved large magnetic field gradients, to find a path for creating non-classical states of motion even though the mechanical periods are measured in milliseconds and the NV coherence times are measured in microseconds. We specifically leverage dynamical decoupling pulse sequences, such as Hahn echo and Carr-Purcell-Meiboom-Gill (CPMG) variations, which are employed to filter out DC fluctuations and access the significantly longer coherence time $T_{2}$ \cite{WangEtAl1, BarGillEtAl, WangEtAl2}, but we also show here that they enable non-classical state generation. Furthermore, we find these same protocols also allow for making highly sensitive, high bandwidth force measurements with 10 s of dB of backaction evasion by only probing the NV spins, rather than any direct observable of the mechanics, building upon prior work \cite{Hsu2016,Slezak2018}. Crucially, the theoretical framework developed here for pulse-driven sensing and entanglement verification is platform-independent, offering a roadmap for high-precision quantum control in diverse hybrid-mechanical systems.

This effort fits into the larger exploration of spin-motion coupling, where the NV center spin is coupled to the motion of an oscillator through various interactions. For example, Rabl et al. \cite{RableEtAl} couple an NV center spin to a resonator's motion through a magnetic field,  Teissier et al. \cite{TessierEtAl} use a strain interaction to couple the spin to the motion of its host diamond cantilever, and Delord et al \cite{DelordNV} couple the spin to the librational modes of the host diamond through a magnetic interaction. 
 
Our work is presented as follows. Section \ref{SetupSection} describes the apparatus; subsection \ref{SetupEW} discusses the generation of entanglement by introducing an entanglement witness (EW), eventually including the effects of a thermal bath, and subsection \ref{SetupForceSensing} details how the setup can be used to sense forces while considering shot noise, thermal noise, and backaction. Section \ref{PulseSequences} considers how to use pulse sequences to implement backaction evasion for improved sensitivity to forces and the consequences this has on the EW. Section \ref{Squeezing} examines how having multiple NV spins can provide an enhancement in the signal-to-noise ratio through squeezing.

\begin{figure*}[htb]
    \centering
    \sidesubfloat[]{
         \includegraphics[width=8.6 cm]{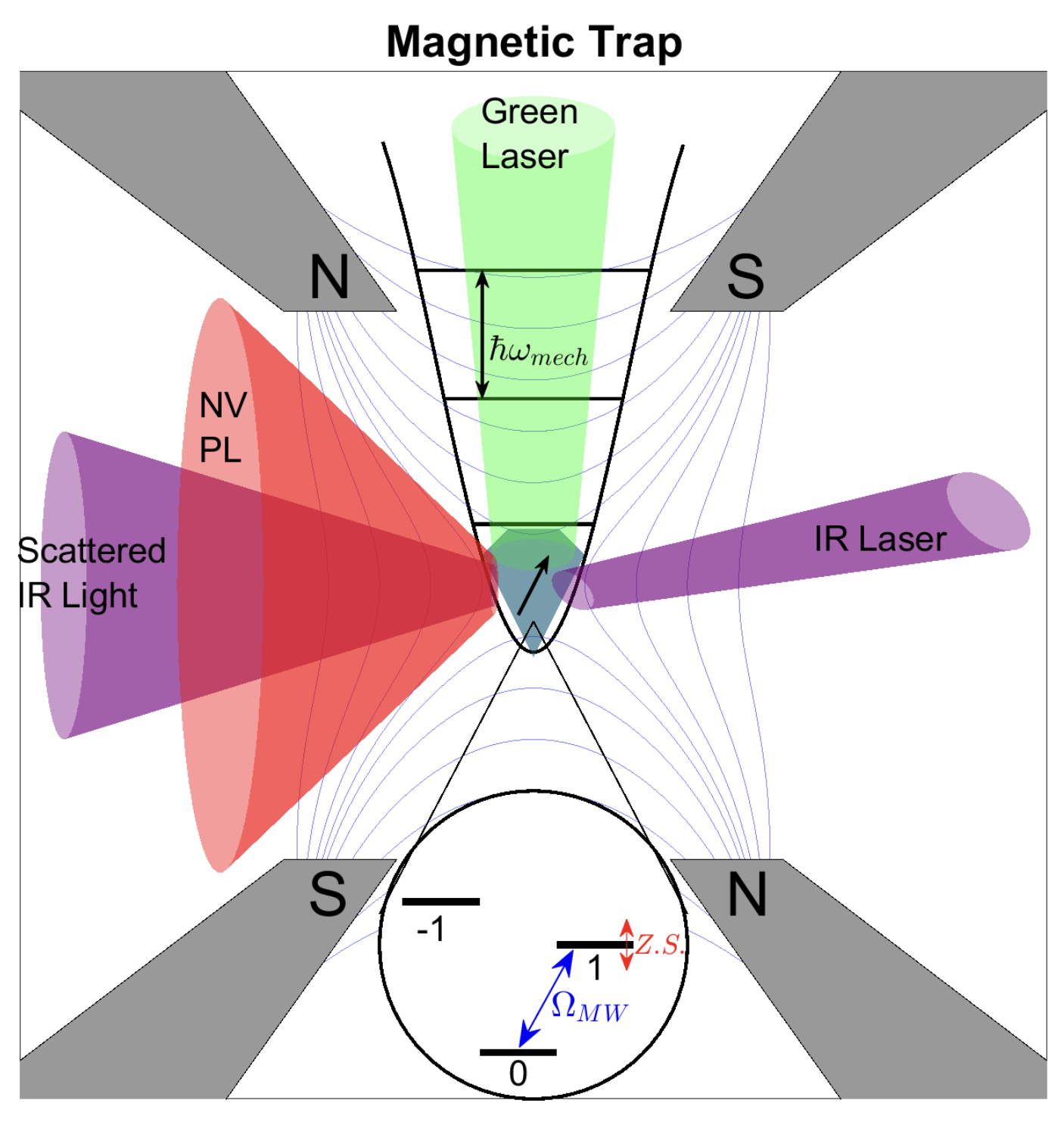}
         \label{Setup}
        }
     \sidesubfloat[]{
         \includegraphics[width=8.4 cm]{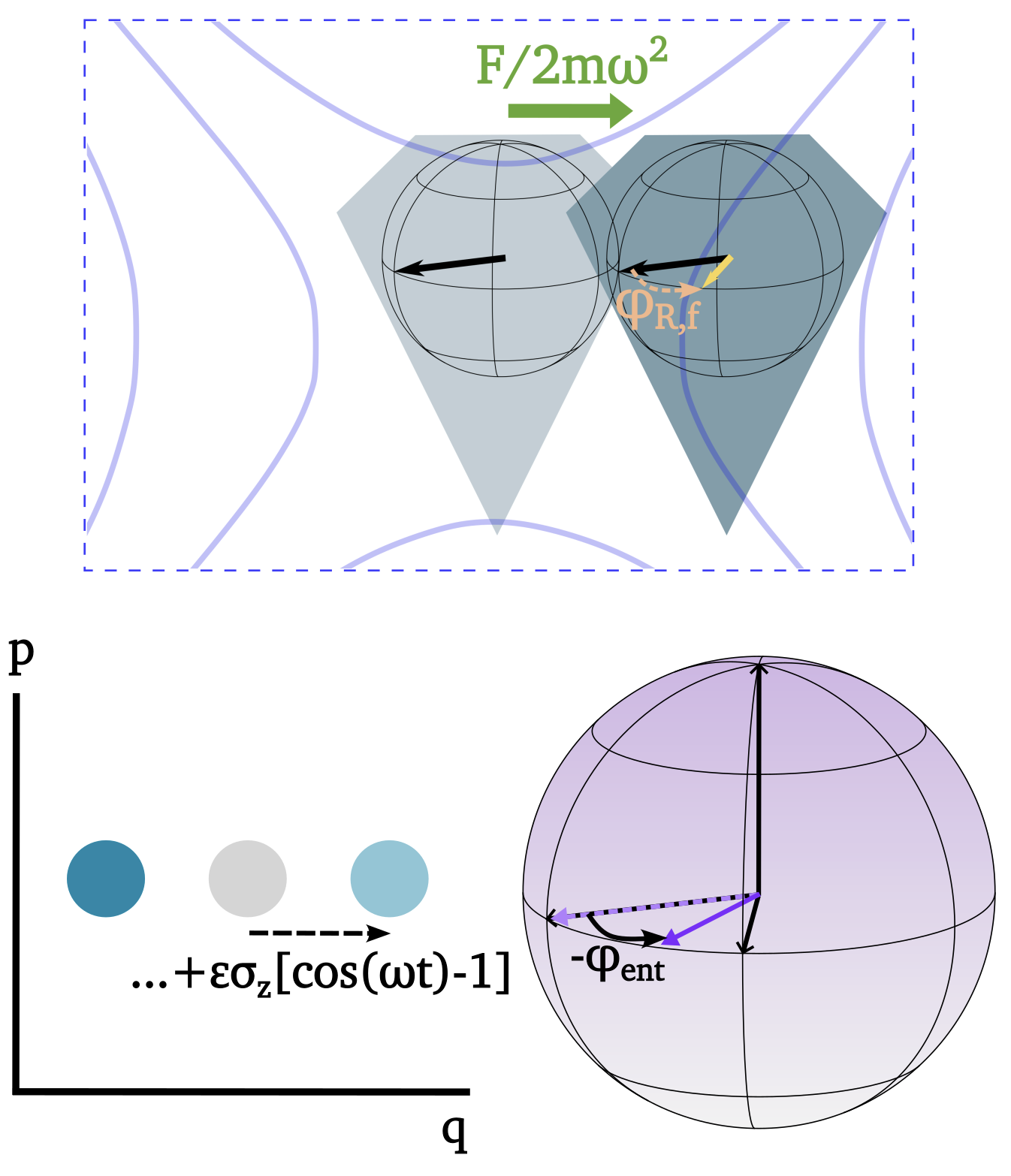}
         \label{HeisOpPic}
        }
    \caption{\textbf{(a)} A diamond (blue) with an NV center spin is levitated in a magnetic trap (magnets shown in gray). A green laser is used for initialization and readout. The photoluminescence emitted by the NV center is shown in red. Purple depicts the infrared (IR) laser used to detect the position of the diamond. The inset shows the three levels of the spin triplet state; the $m_{s}=-1$ and $m_{s}=+1$ state separate due to Zeeman splitting (``Z.S"). In our setup, we make use of only the $m_{s}=0$ and $m_{s}=+1$ states. \textbf{(b)} The \textbf{first row} illustrates an external force $F$  acting on the levitated diamond. The diamond is initially in the potential minimum. The force causes it to move a distance $\sim F/(2m\omega^{2})$, such that it eventually lies outside the potential minimum. The \textbf{second row} is a visualization of entanglement between the NV spin and the diamond, which can be understood by studying operators under Heisenberg evolution. The operators for the pseudo-spin and the oscillator get correlated by the interaction over time, leading to entanglement and the ability to violate the separable bound of the entanglement witness \cite{PremawardhanaEW}. For example, $q$ varies over time with a term proportional to $\sigma_z$, where $\epsilon \sim \sqrt{1/2 m\omega}\lambda$ and $\lambda = 2g/\omega$. The rotation of the pseudo-spin operators, characterized by the angle $\phi_{\rm{ent}}$, has a similar dependency on the oscillator operators. The exact expressions for $p$ and $q$ plots are in Equations \ref{eq:qHeis} and \ref{eq:pHeis}, whereas that for $\phi_{\rm{ent}}$ is as given in Equation \ref{eq:PhiEnt}.}
    \label{fig:Setup}
\end{figure*}

\begin{figure*}[htb]
    \includegraphics[width = \textwidth]   {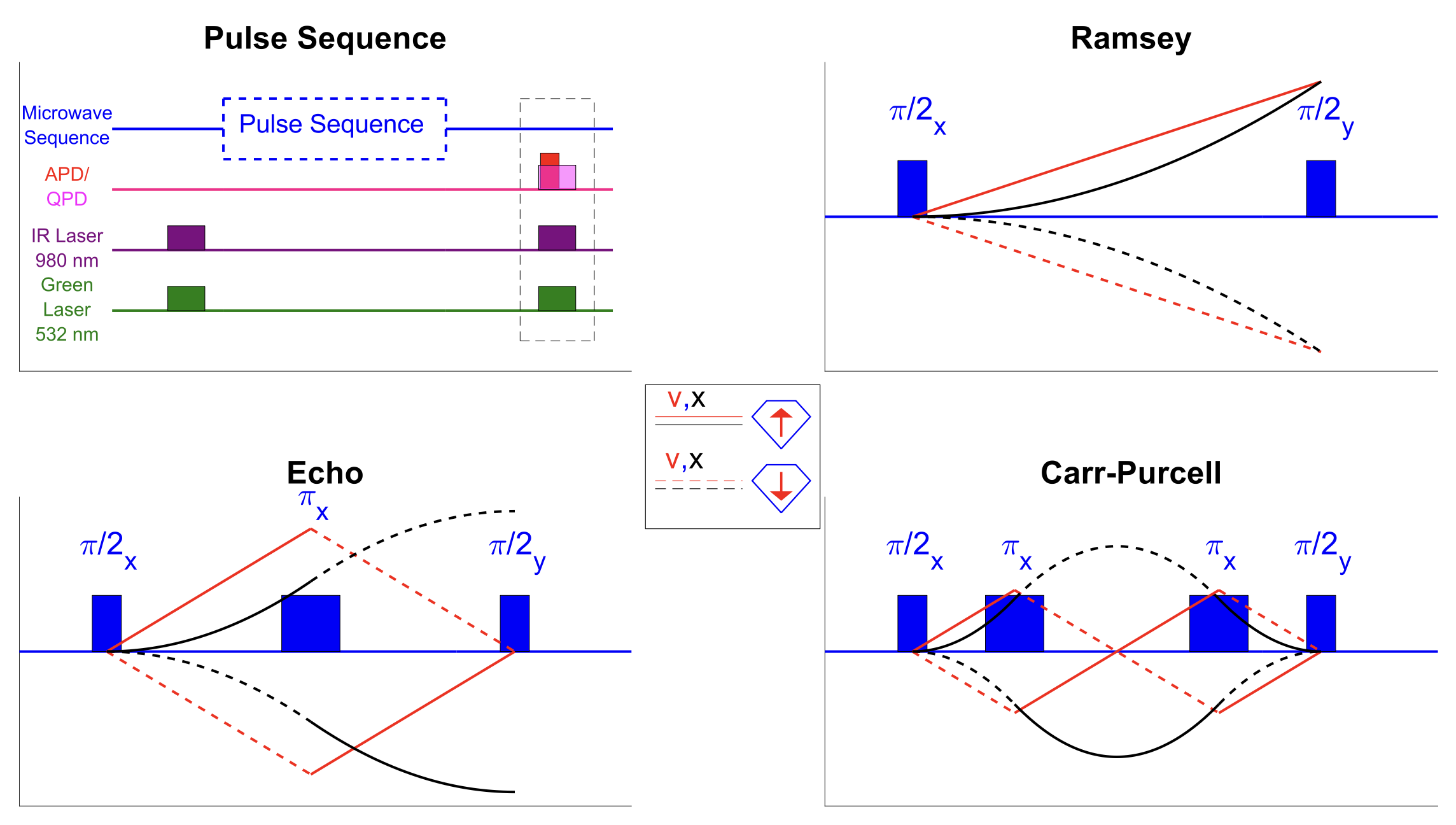}
   \caption{To implement backaction evasion, three pulse sequences, Ramsey, Echo, and Carr-Purcell are investigated. The $v,x$ plots in red and black show the differing levels of success; we see that Ramsey, Echo, and Carr-Purcell have increasingly smaller spin-dependent final mechanical states, leading to reduced backaction.}      \label{fig:PulseSequence}
\end{figure*}

\section{Experimental setup for entanglement and force sensing}
\label{SetupSection}

\figref{fig:Setup}(a) shows the geometry of the magneto-gravitational traps that the proposed experiments will use~\cite{Hsu2016,Slezak2018,peng2018thesis}. Unlike optical or ion traps, which are subject to photon-recoil heating and electric field noise, the magnetic trapping of diamagnetic crystals offers a decoherence-free environment relative to light-matter interactions, potentially enhancing mechanical coherence. The magnetic potential energy for such a particle is given by $U_B = -(V/2) (\mu - \mu_0) \lvert \bvec{H}(x,y,z)\rvert^2$, where  $\mu_0$ is the magnetic permeability of free space, $\mu$ is the magnetic permeability of the particle, and we neglect variations in the field across the particle. Since stable trapping requires $\nabla^2 U > 0 $, only diamagnets ($\mu < \mu_0$) and superconductors ($\mu = 0$) can be stably levitated in static fields.  In particular these particles will be repelled by magnetic poles and attracted to field minima. To make traps in all three dimensions, the magnetic flux from two pieces of permanent magnets is focused by four high-magnetization material pole pieces and configured into a quadrupole distribution. For more details on magnetic levitation, see References~\cite{Hsu2016,peng2018thesis}.

To establish the theoretical framework, we begin with a generalized model of a spin-1 system (here, the NV center) coupled to a single motional degree of freedom ($x$) via a linear magnetic gradient. The effective magnetic field along the NV quantization axis is given by $B(x) = B_0 + \partial B \cdot x$, where $\partial B$ represents the spatial variation of the field. Assuming the mechanical frequency is small relative to the Zeeman splitting, we apply the rotating wave approximation to focus on the longitudinal spin-motion coupling, neglecting rapidly oscillating transverse terms.

Thus, we obtain the toy model Hamiltonian, 
\cite{RableEtAl,BauchEtAl}:
\begin{equation}
H/\hbar = D S_z^2 + \gamma_e B_0 S_z + 2\Omega(t) S_x + 2 g S_z (a + a^\dag) + \omega a^\dag a
\label{eq:NVhamilGen}
\end{equation}
where $D$ is the typical $2 \pi \times 2.88$ GHz splitting of the ground state triplet of the NV center, $\Omega(t)$ is the microwave drive used to generate ESR pulses driving the spin, and the harmonic motion of the crystal is given by $\omega$ with creation operator $a^\dag$. We use natural harmonic oscillator units with $g = \gamma_e \partial B\  x_0 /2$ where $x_0 = \sqrt{\hbar/2m\omega}$ is the harmonic oscillator length, $\omega$ is the harmonic oscillator frequency, and $\gamma_e = g^* \mu_B/\hbar \approx 2 \pi \times 27$ GHz/T is the gyromagnetic response of the NV center spin.

For the rest of this paper, we take $\hbar=1$ unless otherwise indicated, and we work under the assumption that only two internal states of the NV center are leveraged, for example, 0 and 1, and thus define the pseudospin $\sigma_z = \ket{0}\bra{0} - \ket{1}\bra{1}$. This yields a simple two-level model of a spin interacting with a harmonic oscillator:
\begin{equation}
    H_{0-1} = \omega_L \sigma_z/2 + \Omega(t)\sigma_x + g \sigma_z (a + a^\dag) + \omega a^\dag a
\label{eq:NVhamilTwoLevel}
\end{equation}
where $\omega_L = \gamma_e B_0 + D$ is the Larmor precession frequency. We will also consider the scenario in which there are multiple (in that case, $N_s$) NV centers in the same crystal observing the same gradient as occurs in a more heavily doped crystal.

The rest of this section shall cover how this setup can be used to generate entanglement between the spin and the motion of the diamond, in addition to discussing its capabilities as a force sensor. We will analyze these settings within the context of various sources of noises, such as thermal noise and backaction.

\subsection{Entangling the spins and the mass's motion}
\label{SetupEW}

As a matter of principle, this system provides a prototypical system for creating large entangled states, characterized by mechanical superpositions much larger than the harmonic oscillator length $x_0$. Specifically, consider the scenario in which one cools the mechanical center-of-mass motion to its ground state, $\ket{\rm vac}$, while the spin is initialized in the $\sigma_z = 1$ state ($\ket{0}$). Then a $\pi/2$ microwave pulse rotates $\ket{0}$ to $\ket{+} = (\ket{0} + \ket{1})/\sqrt{2}$ \cite{PRXQuantum.2.030330}. Waiting a time $\tau$, we find in the rotating frame the state
\begin{align}
\ket{\Psi} =& \frac{1}{\sqrt{2}} \left(\ket{0} \ket{-\tilde{\alpha}} + \ket{1} \ket{\tilde{\alpha}}\right) \\
\tilde{\alpha} =& - i g \int_0^\tau e^{-i \omega t} dt =  - \frac{g}{\omega} \left[i \sin(\omega \tau) -  (\cos(\omega \tau) - 1) \right]
\end{align}
where $\ket{\tilde{\alpha}}$ is a coherent state.

For a micrometer-sized diamond particle a typical mass is 3 pg with a mechanical resonance of $\omega = 2 \pi \times 100$ Hz in a gradient $\partial B \sim 10$ kT/m. This yields $g/\omega \sim 2$. More generally, 
\begin{equation}
\frac{g}{\omega} \sim 0.3 \left(\frac{dB}{\rm 1\ kT/m} \right) \left(\frac{\rm 1\  pg}{m} \right)^{1/2} \left(\frac{\rm 2 \pi \times 100\  Hz}{\omega} \right)^{3/2}
\end{equation}
Here, we see that $g/\omega$ is of order unity; a `large' entangled state would be readily accessible if both mechanical and spin coherence times are long enough and if ground state cooling is available. Since this quantity is important, we define $\lambda = 2g/\omega$ -- including the factor of 2 here is a choice we make for the upcoming calculations.

In principle, this type of entangled state can also be distinguished using an entanglement witness (EW) -- a statistical test of entanglement particularly well suited to combined spin-and-mechanical systems. An EW has a bound for separable states. If the experimental value of the EW violates this bound, one can conclude that the two systems are entangled. We closely (and sometimes directly) follow the calculations in our previous paper, Premawardhana et al.  \cite{PremawardhanaEW}. As such, the entanglement witness for this setup is as follows:
{\small
\begin{equation}
W = 1/4 \textrm{Var}(\sigma_x)+\textrm{Var}( \sigma_y/2+a_y q + b_y p)+\textrm{Var}(\sigma_z/2 +a_z q + b_z p).
\label{eq:NVwitness}
\end{equation}}
As can be seen, this EW makes use of correlations between the different observables of the two different systems, i.e., the pseudo-spin operators $\sigma_\mu$ which describe the NV spins (here $\sigma_\mu$ is the usual Pauli matrix), and the position and momentum operators, $q$ and $p$, which describe the nanodiamond oscillator. When employing Eq. \ref{eq:NVwitness} to determine the existence of entanglement, it is essentially these correlations that distinguish the experimentally-targeted state from any separable state. In choosing a witness, one has the freedom of choosing $a_\mu$ and $b_\mu$ to maximize the violation observed for the experimentally-targeted state compared to any separable state. 

In the subsequent EW calculations, it is assumed that the NV center is in the $\sigma_x$ eigenstate $\ket{+}=(1/\sqrt{2})( \ket{0}+\ket{1})$ at $t=0$. We also drop the $\Omega(t)\sigma_x$ term in Eq. \ref{eq:NVhamilTwoLevel} in the following evaluation.

$W$ has the following bound for separable states:
\begin{equation}
W_b =\frac{1}{2}+|a_y b_z - a_z b_y|
\label{eq:Wb}
\end{equation}
We must also obtain $W_{\textrm{en}}$, where $W_{\textrm{en}}$ is the value taken by Eq. \ref{eq:NVwitness} when the variances are evaluated for the expected entangled state, in order to find good choices of $a_\mu$ and $b_\mu$. 

For the single-spin case where the oscillator starts in a thermal state, with thermal occupancy $\bar n$, we optimize $a_\mu$ and $b_\mu$. As explained in Ref. \cite{PremawardhanaEW}, we solve (simultaneously) for $a_{\mu}$ and $b_{\mu}$, such that $\partial_{a_{\mu}}W_{\textrm{en}}=0$ and $\partial_{b_{\mu}}W_{\textrm{en}}=0$. At t = $\pi/\omega$ (and $\omega_{L}=0$), which is when maximal entanglement occurs for $\bar n = 0$, these coefficients evaluate to:
\begin{align}
a_y & = 0
\label{eq:ayhalfperiod}\\
b_y & = \sqrt{\frac{2}{m \omega}}\lambda e^{-2(2 \bar n + 1) \lambda^{2}} 
\label{byhalfperiod}\\
a_z & = \sqrt{2 m \omega} \frac{\lambda}{(1+2 \bar n +4 \lambda^{2})}
\label{eq:azhalfperiod}\\
b_z & = 0.
\label{eq:bzhalfperiod}
\end{align}
Next, we calculate the value $W_{\rm{en}}$ that Eq. \ref{eq:NVwitness} takes for the target entangled state, in addition to $W_{b}$ as given in Eq. \ref{eq:Wb}:
{\small
\begin{align}
W_{\rm{b}} & = \frac{1}{2}+ \frac{2 e^{-2(1+2\bar n)\lambda^{2}}\lambda^{2} }{1+2 \bar n +4 \lambda^{2}}
\label{eq:Wbhalfperiod}\\
W_{\rm{en}}&= \frac{1}{2}+\frac{1+2\bar n}{4(1+2 \bar n+4 \lambda^{2})}-\frac{e^{-4(1+2\bar n)\lambda^{2}}}{4}[1+4(1+2\bar n)\lambda^{2}]
\label{eq:Wenhalfperiod}
\end{align}}
As in Ref. \cite{PremawardhanaEW}, the quantity $W_{\rm{ratio}}=(W_{b}-W_{\textrm{en}})/W_{b}$ is of interest. $W_{\rm{ratio}}$ is positive if the EW is violated; furthermore, it informs us of the number of measurements ($n_{\rm{meas}}$), since $n_{\rm{meas}} \sim W_{\rm{ratio}}^{-2}$. The main difference between the EW results in this paper and those of our previous paper (Ref. \cite{PremawardhanaEW}) is as follows: here, we present our results to all orders of $\lambda$ for a single spin, whereas in our previous work, we present to order $\lambda^{2}$  for $N$ spins. This allows us to investigate entanglement for the high coupling present in the nanodiamond system. Note that we do not maximize $W_{\rm{ratio}}$ due to mathematical complexity, as also explained in Ref. \cite{PremawardhanaEW}.

It is interesting to understand how $\bar n$ and $\lambda$ affect $W_{\rm{ratio}}$. In $W_{b}$, $\bar n$ appears as a negative exponential and in the second term's denominator, such that $W_{b} \rightarrow 1/2$ as $\bar n \rightarrow \infty$. This in contrast to the second term of $W_{\rm{en}}$, which has linear $\bar n$ in both the numerator and denominator. This means as $\bar n \rightarrow \infty$, $W_{b}<W_{\rm{en}}$, showing that large $\bar n$ is detrimental to the violation of the witness (since for violation we need $W_{\rm{ratio}}>0$), as expected.

On the other hand, $\lambda$ behaves in a more complex manner. In the second term of $W_{b}$, $\lambda^{2}$ appears in the numerator as both an exponential and a factor, and as a term in the denominator. In $W_{\rm{en}}$, $\lambda^{2}$ appears in the denominator of the second term, and as an exponential and a factor in the third term. Taken together, this indicates that increasing the coupling is advantageous up to a certain point, after which the exponentials take over and we are left with $W_{b}<W_{\rm{en}}$. 

The general expressions for $W_{b}$ and $W_{\rm{en}}$, along with the details of the calculation, are given in Appendix \ref{NoiselessEWapp}.

We now add more detail to this analysis by having  the oscillator be in continuous contact with a white noise bath, as can be described by the Hamiltonian
\begin{equation}
    \tilde H_{0-1} = \omega_L \sigma_z/2 + \Omega(t) \sigma_x+ (g \sigma_z+f_{\textrm{th}}(t)) (a + a^\dag) + \omega a^\dag a
\label{eq:NVhamilTwoLevelfactors}
\end{equation}
with (as also discussed in Appendix \ref{a:magnus})
\begin{equation}
\langle\langle{f_{\rm th}(t) f_{\rm th}(t')}\rangle\rangle = 2 \gamma \bar n \delta(t-t').
\end{equation}
where $\bar n = k_b T/\hbar \omega$.

 \begin{figure}[htb]
    \centering
    \sidesubfloat[]{
         \includegraphics[width=8 cm]{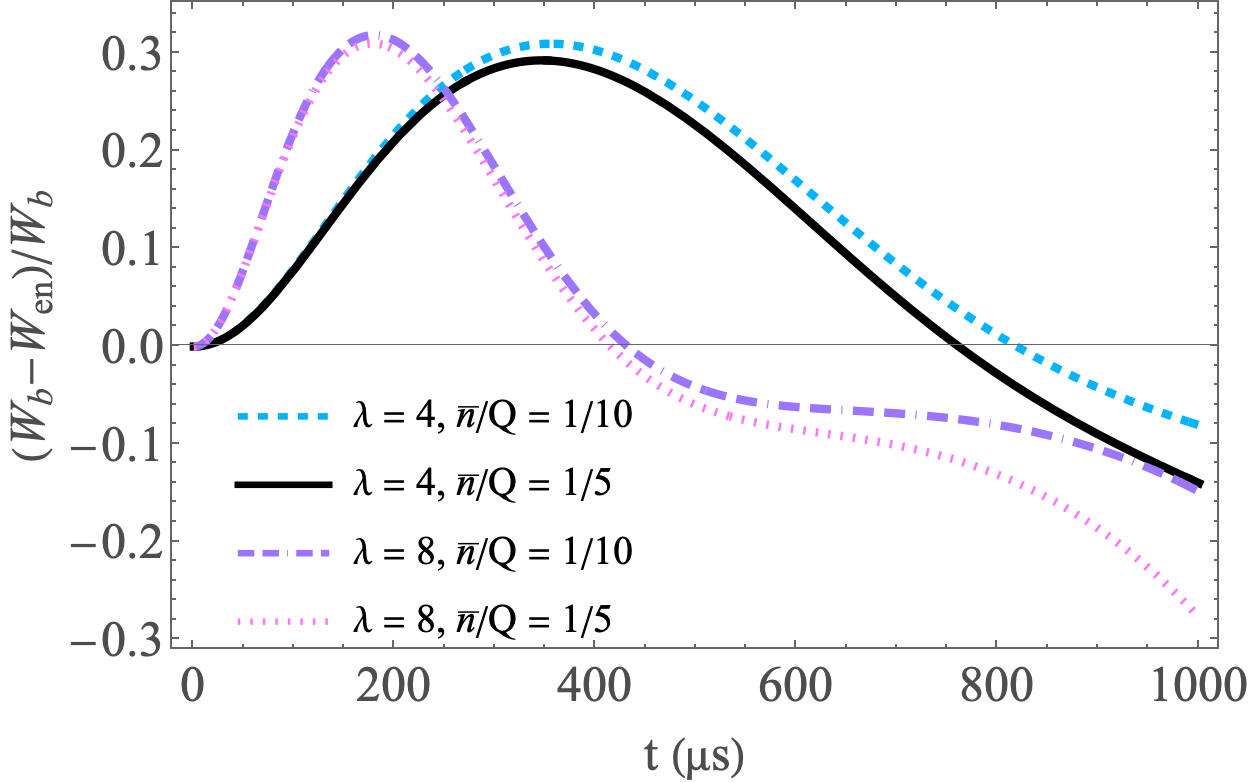}
         \label{fig:GroundStatePlusBath}
        }
    \hfill
    \sidesubfloat[]{
        \includegraphics[width=8 cm]{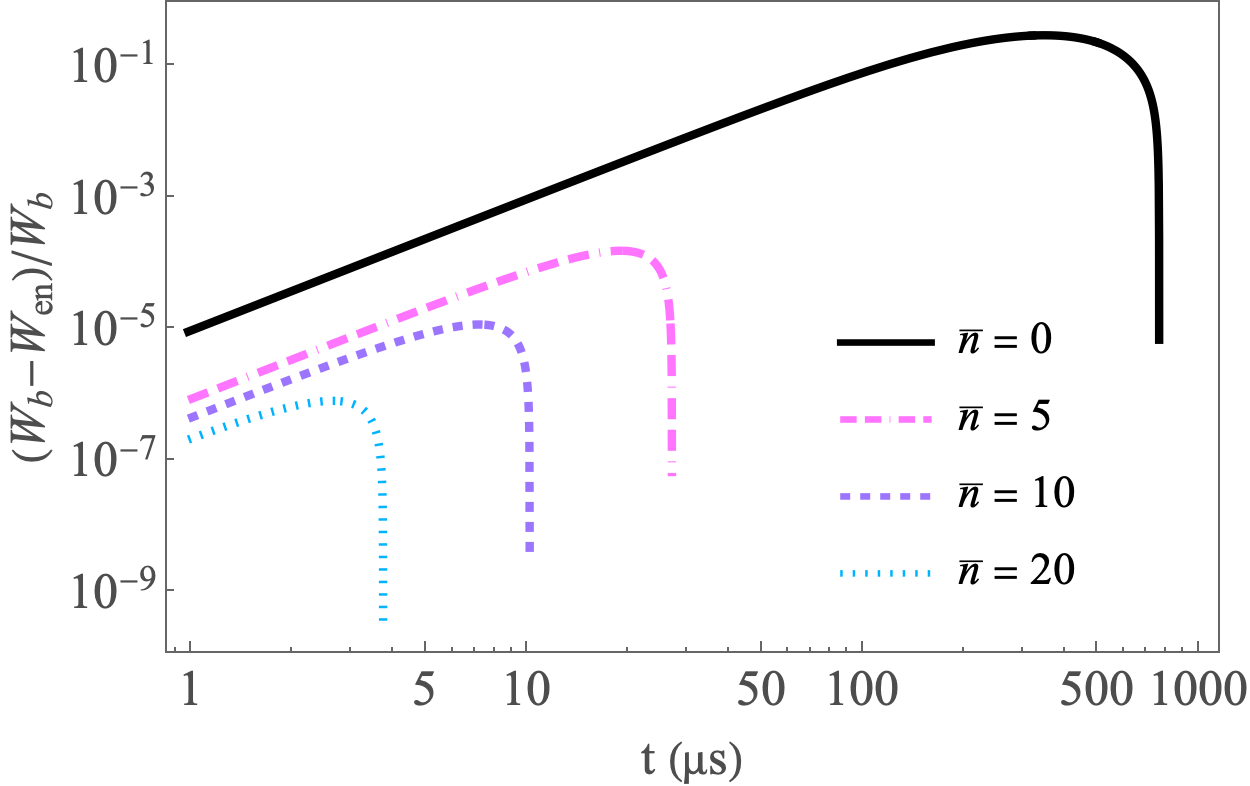}
        \label{fig:ThermStatePlusBath}
      }
    \caption{In these figures, $\omega = 2 \pi \times 100$ Hz. (a) Plot of $(W_{b}-W_{\textrm{en}})/W_{b}$ vs. time for the case where the mass is cooled to its ground state at the start of the experiment and remains in contact with a continuous thermal bath. With higher $\lambda$, maximum violation is reached faster. The lower the value of $\bar n/Q$, the higher the violation. (b) Plot of $(W_{b}-W_{\textrm{en}})/W_{b}$ vs. time for the case where the mass starts in an initial thermal state and remains in contact with a continuous thermal bath. For all curves, $\lambda = 4$ and $\bar n/Q = 1/5$. The plots get truncated when the violation ceases to exist. Increasing $\bar n$ decreases the time for which violation exists.}
    \label{fig:EWbathPlots}
\end{figure}
$W_{b}$ and $W_{\rm{en}}$ must now be re-evaluated. Since the expressions are rather complex, we present plots for different scenarios in Figure \ref{fig:EWbathPlots}. Figure \ref{fig:GroundStatePlusBath} shows $(W_{b}-W_{\textrm{en}})/W_{b}$ against time for different combinations of $\lambda$ and $\bar n/Q$ for when the oscillator starts in the ground state. A higher $\lambda$, and thus, a higher coupling, results in maximal violation being reached faster. Furthermore, a lower value of $\bar n/Q$ results in a higher value for this maximal violation, though, arguably, for the parameters explored here, the difference is only slight. Next, in Fig. \ref{fig:ThermStatePlusBath}, we explore the effects of starting in a ground state vs a thermal state. For $\lambda = 4$ and $\bar n/Q = 1/5$, we plot $(W_{b}-W_{\textrm{en}})/W_{b}$ against time for different values of $\bar n$. When $\bar n = 0$, the oscillator starts in the ground state. The effect of starting in a thermal state is very apparent; it significantly deteriorates the duration for which violation exists. Indeed, comparing both figures, increasing the thermal occupancy of the initial thermal state $\bar n$ affects the violation much more than the same factor of increase in the thermal bath parameter $\bar n/Q$. When calculating the violation for the case where the oscillator starts in the ground state, we used the coefficients $a_{\mu}$ and $b_{\mu}$ that correspond to the scenario where $\bar n = 0$ at the start (despite the fact that there is a thermal bath, since that is separately characterized by $\bar n/Q)$, while the thermal state coefficients were used when starting in a thermal state.

Further details of this calculation are provided in Appendix \ref{ThermalBathApp}.

In reality, applying the EW in  Eq. \ref{eq:NVwitness} has to overcome the issue of the NV spin having very short coherence times. Indeed, to have coherence times of even 300 $\mu$s, and to also implement back-action evasion, pulse sequences have to be applied. These pulse sequences can result in the decoupling of the spin from the momentum, thus potentially decreasing the effectiveness of this EW, since it relies on correlations between the spin operators and momentum. In Section \ref{EWwithPulses}, we find that Eq. \ref{eq:NVwitness} can still be used; however, the exact expressions for $W_{b}$ and $W_{\rm{en}}$ must be modified.

\subsection{Force sensing with the entangled state}
\label{SetupForceSensing}

Extensive analysis has been done on the sensing potential of a variety of levitated nanoparticle and trapped ion systems, especially after taking into account various sources of noise, such as thermal noise and backaction, among others \cite{Yoctonewtonforcedetection,Hempston2017,Mason2019,PhysRevLett.125.181102,BiercukEtAl,IvanovEtAl}. This section details such an analysis for the setup discussed in this paper.

Our levitated nanodiamond system has a large spin-motion coupling, with $\lambda = 2 g/\omega > 1$. At the same time, NV centers have a long history of use as spin-based sensors of magnetic fields. While levitated particles have a remarkably `quiet' environment (represented by extremely high quality factor $Q = \omega/\gamma$), monitoring their position to the so-called standard quantum limit usually comes with a number of tradeoffs, particularly in optical power and measurement apparatus. Here we show how we can monitor the motion using only the spin degrees of freedom.

Our particular focus will be on force and impulse sensing, taking advantage of the small mass of the diamond. Such forces could be applied by external electric or magnetic fields, or other sources, e.g., radioactive decay in the crystal leading to recoil. To motivate the expected sources of noise and imperfection, we start with the standard approach for magnetic sensing using a single NV spin. Specifically, the sensitivity for detecting a magnetic field that induces a frequency shift of the NV center's magnetic resonance $\delta = \gamma_e \delta B$ using Ramsey spectroscopy (see Fig.~\ref{fig:PulseSequence}) is given by 
\begin{equation}
    \eta_\delta = \frac{1}{C} \frac{1}{\gamma_e \sqrt{T_2^*}} \label{e:sens}
\end{equation}
where $C \sim 0.1 - 0.3$ represents the combination of duty cycle and collection (in)efficiency. The inhomogeneous broadening dephasing time $T_2^*$ enters due to the lack of an echo sequence to remove low frequency noise sources, including paramagnetic and nuclear spins in the crystal. The main sources of noise in this analysis are the inhomogeneous broadening and spin projection noise -- each measurement yields only a single bit of information, leading to Equation~\ref{e:sens}.

Next we analyze how this experimental setup can be used as a force sensor by adding an external force to the Hamiltonian,
\begin{equation}
V = -f (a + a^\dag),
\end{equation}
where $f = F(t) x_0/\hbar$ is in general a time-dependent force, which includes both a signal field and thermal heating effects, as described in Appendix~\ref{a:magnus}. The influence of external forces can be detected by monitoring the NV center spin state, where the motion transduces to a change in the magnetic resonance frequency -- a technique previously demonstrated in Ref.~\cite{NeukirchEtAl2}. 

Let us return to the magnetic sensitivity, and ask: what is the force sensivity we would expect if that were the only limit?
For low frequency forces, we expect the harmonic response will lead to a shift of the average center of mass of the diamond by an amount $\sim f / \omega$ when integrated for times $t$ similar to or longer than the mechanical period. This in turn leads to a mean shift in the Larmor frequency of $g f / \omega$. In physical units, $g \frac{f}{\omega} = \frac{\gamma_e \partial B}{\omega}  F x_0^2/\hbar = \gamma_e \partial B\ \frac{F}{2 m \omega^2}$, which is exactly the shift of the Larmor frequency due to the shift of the equilibrium position of the mass from the incident force, as expected.

Thus, as a baseline of expected sensitivity to forces for this type of system, one would hope to be able to achieve at least a sensitivity similar to that of the magnetic sensing of the sensor to this shift, i.e., a force sensitivity of: 
\begin{equation}
\eta \sim \frac{\omega}{g \sqrt{T_2^*}}\ {\rm H.O.\ units} = \frac{2 m \omega^2}{\gamma_e \partial B \sqrt{T_2^*}} \ {\rm SI\ units}
\label{eq:eta1}
\end{equation}
where $T_2^*$ is the inhomogeneous dephasing that we would naively observe with a Ramsey sequence to estimate the resonance frequency. 
For a 3 pg diamond with a frequency of $2 \pi \times 100$ Hz and a gradient of 10 kT/m, \eqnref{eq:eta1} yields a sensitivity of $1.4 \times 10^{-21} \text{ N}/\sqrt{\rm Hz}$.  

This limit can be improved by having more than one NV center contributing to the signal. Using $N_s$ contributing spins further improves the sensitivity by $\sqrt{N_s}$. For typical micrometer-sized crystals, we expect $N_s \lesssim 10^6$ for a doped crystal (1-10 ppm), and $\sim 1$ for a lightly doped sample.

\textbf{Thermal noise considerations:}
The next practical consideration when building such a force sensor arises from the \textit{thermal noise incident on the mechanical system}. For a typical Brownian motion bath that describes high Q mechanical system noise when $k_b T \gg \hbar \omega$, we expect that the force term includes thermal noise whose variance grows as $\gamma \frac{k_b T}{\hbar \omega} t$ for an integration time $t$. This produces a limit to the sensitivity that goes as 
\begin{equation}
\eta_{\rm th} \sim \frac{\hbar}{x_0} \sqrt{ \gamma \frac{k_b T}{\hbar \omega} }
\end{equation} 
in units of N/$\sqrt{\rm Hz}$. At room temperature for our 3 pg, $2 \pi \times 100$ Hz system, this gives $\eta_{\rm th} \approx 1.3 \times 10^{-16} \frac{1}{\sqrt{Q}} ( T / 300\ {\rm K})$ N/$\sqrt{\rm Hz}$ where $Q = \omega / \gamma$ is the mechanical quality factor. For typical quality factors $Q \gtrsim 10^6$, the thermal noise dominates over the shot noise limit. 

As an aside: at higher Q $\sim 10^7$ and larger mass (3 ng) this is no longer true: now the shot noise dominates over the thermal noise. This also occurs with higher frequency mechanical modes, e.g., those at $2 \pi \times 1$ kHz and higher. 

\textbf{Backaction noise limit:} In addition to the thermal noise and the shot noise (spin projection noise), we may eventually be limited by the so-called \textit{standard quantum limit}. Here, the spin-dependent force on the mass of the diamond leads to backaction, which looks like heating of the center of mass through the measurements -- this arises simply from the uncertainty of the spin-dependent motion post measurement. A simple calculation shows that this can be a point of concern, following the pedagogical approach of Caves \cite{RevModPhys.52.341}. 

First, the spin measurement of $N_{s}$ spins infers the position variable via the frequency shift of the spins; for a spin free precession time of $\tau \lesssim T_2^*$, we expect an imprecision in position $\Delta x \sim [\gamma_e \partial B \tau \sqrt{N_s}]^{-1}$ due to the unit spin shot noise. This imprecision leads to an uncertainty in the momentum via the Heisenberg uncertainty relationship $\Delta p \sim \hbar / \Delta x$. If the next measurement occurs a time $\sim t \geq \tau$ later, we can attempt to balance the backaction-induced uncertainty, $t \Delta p/m$, with the measurement imprecision, $\Delta x$. This is optimized for $t \hbar/m \Delta x = \Delta x$ which gives the SQL and corresponding optimal magnetic gradient:
\begin{align}
\Delta x_{\rm SQL} &= \sqrt{\frac{\hbar t}{m}} \ ,\\
\gamma_e \partial B^* \sqrt{N_s} &= \frac{1}{\tau \Delta x_{\rm SQL}}  \ .
\end{align}
For a micrometer-sized diamond particle of mass 3 pg with even a single NV spin, taking $t \sim 2 \tau \sim 4 T_2^*$ yields the SQL at $\partial B \gtrsim 9$ kT/m. The corresponding force sensitivity at this repetition rate of measurements and magnetic gradient is $3 \times 10^{-18}$ N/$\sqrt{\rm Hz}$ -- and this only gets worse with more NV center spins. We remark that this noise is larger than the thermal noise for $Q \gtrsim 10^6$.

Looking at these three sources of noise, we see that the short time between measurements and strong spin-motion coupling lead to the curious scenario where back action is a real limit to the use of NV centers as force sensors, even though shot noise is not a limiter when compared to thermal noise. This arises due to both the strong coupling strength and the short time between measurements $\tau \ll 1/\omega$.

\section{Backaction evasion and improved entanglement with pulse sequences}
\label{PulseSequences}

As discussed qualitatively above, the large $g/\omega$ leads to substantial backaction within the short (faster than the mechanical period) coherence times provided by NV centers, which in turn becomes a limit to the performance of the force sensor. Short coherence times also suggest that the entanglement witness described above will not be realizable in practice, since long coherence times are necessary for an entanglement protocol. 

A possible path forward is as follows: increase the spin coherence using simple echo, or more complex Carr-Purcell, sequences. We also find that this will reduce the backaction and enable us to find a balance between the larger $g/\omega$ and the effects of decoherence and heating. As we will also see, the use of pulses will mean that the entanglement analysis discussed previously must be adjusted.

In order to understand why pulse sequences are required, we shall now consider the specifics of measuring the spin precession due to the incoming force. As different pulse sequences are considered, in the interaction picture we will take the coupling $g$ to be time dependent, flipping sign every time a $\pi$ pulse is applied to the spin. Here, we do the calculation for the free evolution period, with $g(t)$ taking into account whether a pulse has been applied or not. We see in the Heisenberg picture that we get Larmor precession for a time $\tau$ during Ramsey interferometry that yields a rotation angle around the $z$ axis of 
\begin{align}
\phi_{\rm R} & = -\omega_L \tau -\int_0^\tau g(t) {\rm Re}[a(t)] dt 
\label{eq:PhiR}\\
a(t) &= a_0 e^{-i \omega t} - i \int_0^t e^{-i \omega(t - t')} [ g(t') S_z-f(t')] dt' 
\label{eq:ForceSensingLadderHeisOp}
\end{align}
where we recognize that for the free evolution period $S_z$ is a constant of motion.

From these equations, we see that the rotation of the spin (its phase) due to the center of mass motion has two terms. One is a contribution from the spin itself, which leads to spin squeezing, whose amplitude for constant $g$ (Ramsey) is given by $\frac{2 g^2}{\omega^2} (\cos(\omega \tau) - 1)$. As an aside, choosing $\tau \omega = 2 \pi n$ for $n \in \mathbb{Z}$ eliminates this effect, but we are interested in $\tau \omega < 1$; thus we instead will consider the squeezing as a resource in the following section. 

The second term is best understood by taking the Fourier transform of the incoming force $f(t) = \int_{-\infty}^{\infty} f(\nu) e^{-i \nu t} \frac{d\nu}{\sqrt{2 \pi}}$. This gives for constant $g$ (Ramsey):
\begin{multline}
\phi_{R,f} = \int f(\nu) \frac{2 (\nu [\cos(\omega \tau)-1] - \omega  [\cos(\nu \tau)-1])}{\nu \omega (\nu - \omega)} \frac{d\nu}{\sqrt{2 \pi}}\\ \equiv \int \chi(\nu) f(\nu) d\nu
\end{multline}
It is convenient to rewrite the spin phase using a response function $\chi(\nu)$, defined by the above.
When $\tau = 2 \pi/\omega$, $\chi(\nu)$ is maximal for $\nu = \omega/2$, with value $-16 g/\omega^2$. Intuitively, the spin accumulates a phase from the force that goes as $g$ times the displacement due to the force integrated over the period $\sim f / \omega \times \tau \sim f/\omega^2$. More details about this calculation are given in Appendix \ref{a:magnus}, including Equation \ref{eq:Phi2}. Critically, in that analysis we generalize the simple Ramsey case to the case of piecewise changes in $g(t)$ in the interaction picture, corresponding to $\pi$ pulses.

Curiously, we note that there is nominally no backaction on the mass from the measurement of this force. To see this explicitly, we can look at the spin-dependent displacement $\frac{1}{2}\partial_{S_z} a(t)$ which is
\begin{equation}
\sqrt{\Delta n} = \frac{1}{2} \left| \int_0^\tau e^{-i \omega(\tau - t')} g(t') dt' \right|
\end{equation}
where we look at the induced displacement from the spin in terms of how many phonons are excited, $\Delta n$. For Ramsey interferometry, we find $\Delta n  = \frac{4 g^{2} \sin^2(\omega \tau/2)}{\omega^2}$. Specifically, the spin-dependent force term integrates to zero for $\omega \tau = 2 \pi n$. However, in practice, this is difficult to achieve while maintaining a good `duty cycle'; we would like $T_2^* \omega \sim 1$  to reduce backaction, but levitated trap frequencies are generally quite slow, e.g., tens of Hz compared to $T_2^* \sim 1-10\ \mu$s. Thus, to be in the regime of reduced backaction, we want to both consider using spin echo techniques to improve spin coherence and look at measuring forces with bandwidths greater than $\omega$, which we call broadband sensing.

The large $g/\omega$ afforded by our setup is favorable for the generation of entanglement. However, our entanglement detection protocol consists of the EW given in Eq. \ref{eq:NVwitness}, which relies on measuring correlations between spin and momentum. Employing pulse sequences can destroy such correlations, as can be seen in Figure \ref{fig:PulseSequence}. Therefore, we must confirm that the EW can still be used.

\subsection{Broadband sensing with pulse sequences}

It is natural to consider a Hahn echo \cite{HahnEcho1} or Carr-Purcell-type sequence \cite{CarrPurcellSeq1} to improve the spin coherence towards $T_2$. This comes at the cost of a reduced response to the incoming force. However, as we will show, the total impact is minimal for broadband or high frequency signals.

\begin{table*}[ht]
\caption{
The spin accumulation $\phi$ normalized by the coupling $g$ and unit incoming force $f$, compared against the heating due to spin projection noise in units of harmonic oscillator quanta divided by the coupling $g^2$. All of these are calculated for signals with $|\nu| < 2 \pi/\tau$, and have total sequence times of $\tau$.
\label{t:comp}}
\begin{center}
\begin{tabular}{|l|c|c|c|c|}
\hline
Seq. & $\phi/(g f)$ &  	$\Delta n/g^2$    &  Force SQL$/\xi^{1/4}$    & SQL coupling $g^*$ ($\xi=1/4$)\\
\hline
Ramsey &		$\frac{\omega \tau^3}{6}$ &	$\tau^2$ 	& $\frac{6 \tau}{\omega \tau^3} = \frac{6}{\omega \tau^2}$	&	$\frac{1}{\tau \sqrt{N}}$ \\
Hahn echo &	$\frac{\omega \tau^3}{8}$ &	$\frac{\omega^2 \tau^4}{16}$  &   $\frac{8 \omega \tau^2}{4 \omega \tau^3} = \frac{2}{\tau}$   &	$\frac{4}{\omega \tau^2 \sqrt{N}}$\\
Carr-Purcell &	$\frac{\omega \tau^3}{32}$ &	$\frac{\omega^4 \tau^6}{1024}$ &  $\frac{32 \omega^2 \tau^3}{32 \omega \tau^2} = \omega$ &	$\frac{32}{\omega^2 \tau^3 \sqrt{N}}$ \\
\hline
\end{tabular}
\end{center}
\label{default}
\end{table*}%

 \begin{figure*}[htb]
    \centering
    \sidesubfloat[]{
         \includegraphics[width=8.2 cm]{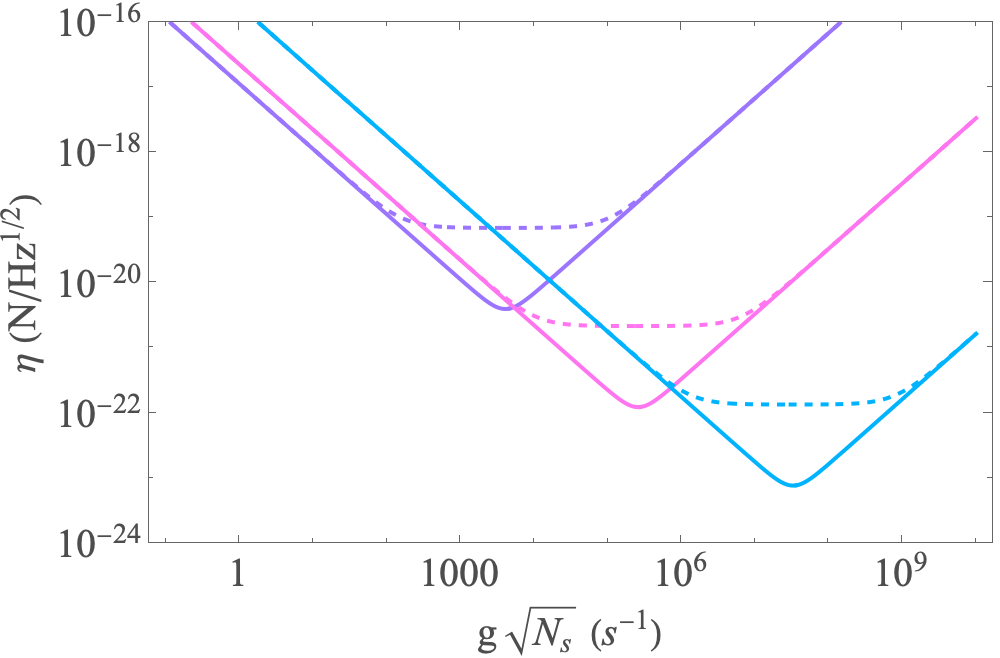}
         \label{fig:SensitivityVSg}
        }
     \sidesubfloat[]{
         \includegraphics[width=8.2 cm]{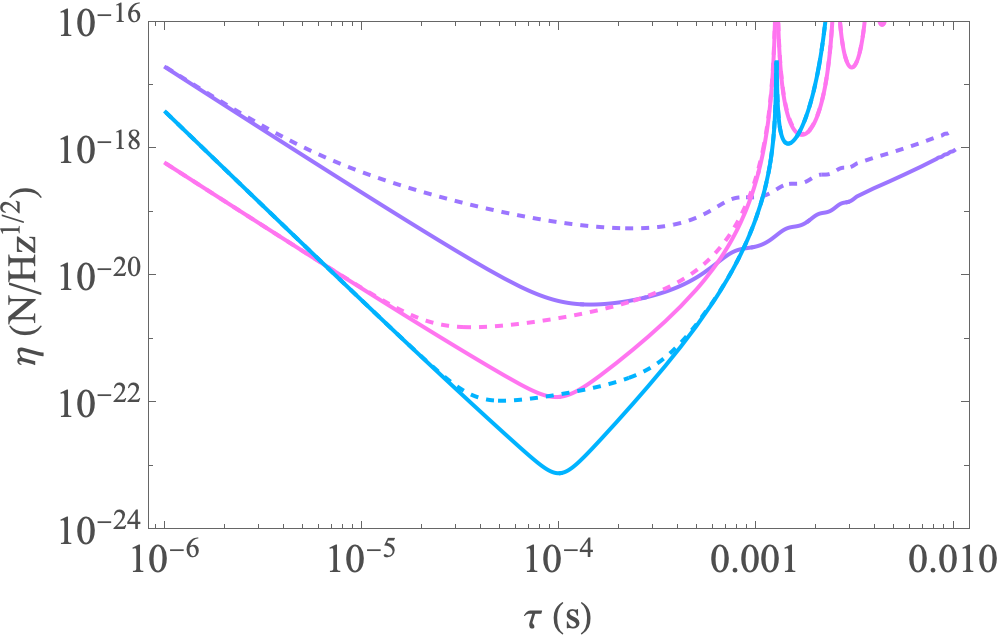}
         \label{fig:SensitivityVStau}
        }
     \hfill
     \sidesubfloat[]{
        \includegraphics[width=12 cm]{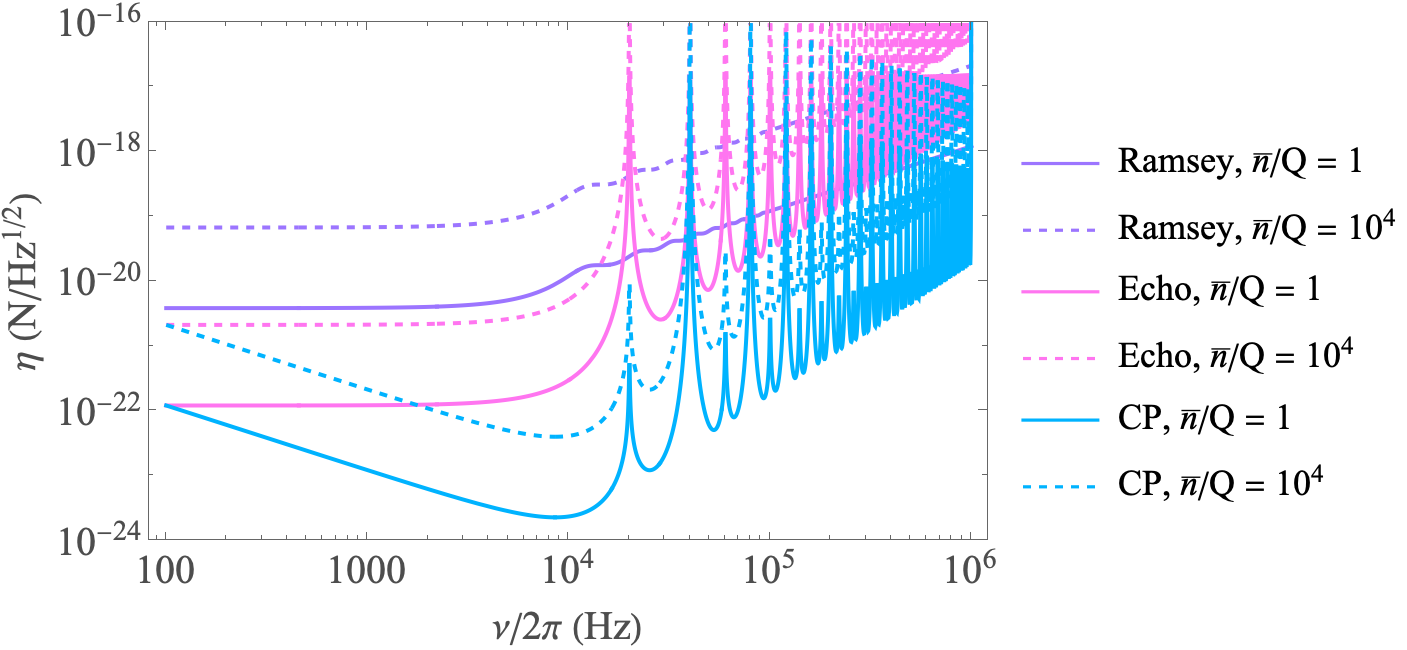}
      \label{fig:SensitivityVSNu} 
      }
    \caption{Plots of sensitivity for force sensing. Different colors indicate the various types of pulse sequences as given in the legend, while solid lines are for $\bar n/Q = 1$ and dashed lines are for $\bar n/Q = 10^{4}$. (a) 1 $\mu$m radius diamond, 100 Hz trap, $Q=10^{6}$, cooling time $t_c$ = 100 $\mu$s, cooling rate $\gamma_c =$ 1 kHz, and an expected signal frequency of 2 kHz. We see the SQL for each type of sequence requires increasing effective coupling to be achieved; improvements of around a factor of 300 are possible in terms of back action evasion. (b)  Optimal g, fixed input of 2 kHz, and varying the pulse sequence length $\tau$ indicates that optimal choice of $\tau$ is near the period of the incoming signal, as expected (c) Optimal $g$ for a nominal 2 kHz signal, fixed time $\tau=$ 100 $\mu$s; as the actual signal frequency varies, the system is responsive in a broad range of frequencies. The characteristic $1/\nu$ behavior of backaction evasion is seen only for the case of Carr-Purcell sequences, consistent with the qualitative understanding that this sequence removes both $x$ and $p$ backaction.}
    \label{fig:SensitivityPlots}
\end{figure*}

We can now ask what happens to (1) the spin and (2) the harmonic motion when we do a Ramsey, spin-echo, or Carr-Purcell sequence on the electron spin during evolution. Of particular interest will be a variant of the Carr-Purcell sequence in which we do two $\pi$ pulses, as follows: free evolve for $t_1$, then $\pi$, then free evolve for $t_1 + t_2$, then $\pi$, then a final evolution for $t_2$. The total time $\tau \approx 2 t_1 + 2 t_2$, and we see that DC spin noise leading to $T_2^*$ is completely removed. At the same time, we do get a small amount of backaction. For $t_1 \omega, t_2 \omega \ll 1$, we find that setting $t_1 = t_2 = \tau/4$ leaves behind a total impulse of
\begin{equation}
\Delta n_{\rm CP} = \frac{g^2}{\omega^2} 2^6 \sin^4(\omega \tau/8) \sin^2(\omega \tau/4) \sim 2^{-10} \omega^4 g^2 \tau^6
\end{equation}
and that this is optimal for $\omega \tau \ll 1$. 

At the same time, we accumulate a force signal from DC to $\sim 2 \pi/\tau$ that is approximated by
\begin{equation}
\phi_{\rm CP} \approx \frac{g \omega \tau^3}{2^5}  \int_{-\infty}^{\infty} f(\nu) e^{-i \nu \tau/2}  e^{- (9 \omega^2 + \nu^2) \tau^2/2^6} \frac{d\nu}{\sqrt{2 \pi}}
\label{eq:PhiCP}
\end{equation}
for $\nu \tau \lesssim 2 \pi$ and $\omega \tau \ll 1$. We plot this approximation
and the actual spectral response function in Fig. \ref{fig:SpectralFunctionPlot} of Appendix \ref{a:magnus}). 

The relative performance of the three sensing modalities—Ramsey, Hahn echo, and Carr-Purcell—is evaluated for a fixed measurement bandwidth $\sim 2 \pi/\tau$. In the low-frequency regime where $\omega \tau \ll 1$, the sensitivities of these sequences are compared in Table~\ref{t:comp}.

Of specific interest is the notional force SQL for each approach to sensing. This arises when considering a sequence of measurements, where the occupation of the harmonic oscillator mode with phonons leaves behind a signal which is comparable to the heating, e.g., a phase noise in the signal that goes as $\sqrt{\Delta n}$. If there is a short cooling period $t_c$ with cooling rate $\gamma_c$, this gives a steady-state occupation of the harmonic oscillator purely from backaction of
\begin{equation}
n^* = \xi \Delta n
\end{equation}
where $\xi \equiv \frac{e^{-\gamma_c t_c}}{1 - e^{-\gamma_c t_c}}$ gives the effect of cooling over many cycles. When we include this term in the noise-to-signal calculation, we have for the noise squared $1/4 + (\Delta n) n^*$, where the first term comes from spin projection noise and the second term from the backaction-induced heating which includes a short cooling period. The noise-to-signal ratio is then
\begin{equation}
\frac{1/4 + (\Delta n)^2 \xi}{\phi^2} \label{eq:nsr}
\end{equation}
As $\phi \sim g$ and $\Delta n \sim g^2$, we find a SQL-like behavior for force sensing at an optimal $g$ such that $1/4 = (\Delta n)^2 \xi$, which gives as the force SQL $\sim \sqrt{\sqrt{\xi} \Delta n} / |\phi/f|$ which is, as expected, $g$-independent.  Comparing the different protocols, we see that the Carr-Purcell sequence well outperforms (has the lowest force SQL) compared to the other sequences for $\omega \tau \ll 1$.

To evaluate the system’s performance, we conduct a numerical study using realistic experimental parameters: a 1 $\mu$m diamond crystal situated in a 100 Hz magnetic trap. Our model incorporates several practical constraints, including a finite experimental duty cycle due to 100 $\mu$s of reset and cooling between measurements, and a finite cooling rate $\kappa$ of 1 kHz. We also consider two different $\bar{n}/Q$ values: $1$ and $10^4$. In the subsequent analysis, we work near a bandwidth comparable to $1/T_2 \sim 10$ kHz, since a typical signal of interest is as such.

The influence of external forces is detected by monitoring the NV center spin state, a technique previously established in the literature \cite{NeukirchEtAl2}. We analyze the sensitivity of this detection channel as follows.

The optimization process begins by determining the ideal effective coupling strength, $g \sqrt{N_s}$, as illustrated in Fig. \ref{fig:SensitivityVSg}. As the coupling gets larger, the more complex sequences start to outperform the simpler sequences due to the reduced dependence on the thermal background. This leads to a higher $g$ before backaction effects.

Upon determining the optimal effective coupling, we investigate the impact of the free evolution period $\tau$ on measurement precision. As shown in Fig. \ref{fig:SensitivityVStau}, $\tau \nu \sim 1$ appears optimal.

Finally, by fixing $\tau$ and sweeping the signal frequency $\nu$, we characterize the spectral response of each sensing modality (Fig. \ref{fig:SensitivityVSNu}). Our analysis reveals distinct operational regimes for each sequence. For both Ramsey and echo sequences, we have a flat response at low frequencies -- they operate as broadband force sensors and are sensitive to DC forces. In contrast, the Carr-Purcell sequence has less sensitivity at lower frequencies, consistent with the usual behavior of a back-action evading impulse detector. Thus for impulse metrology it is likely the optimal approach when compared to the other options.

\subsection{Entanglement witness with pulse sequences}
\label{EWwithPulses}

\begin{figure}
    \centering
    \includegraphics[width=8.6 cm]{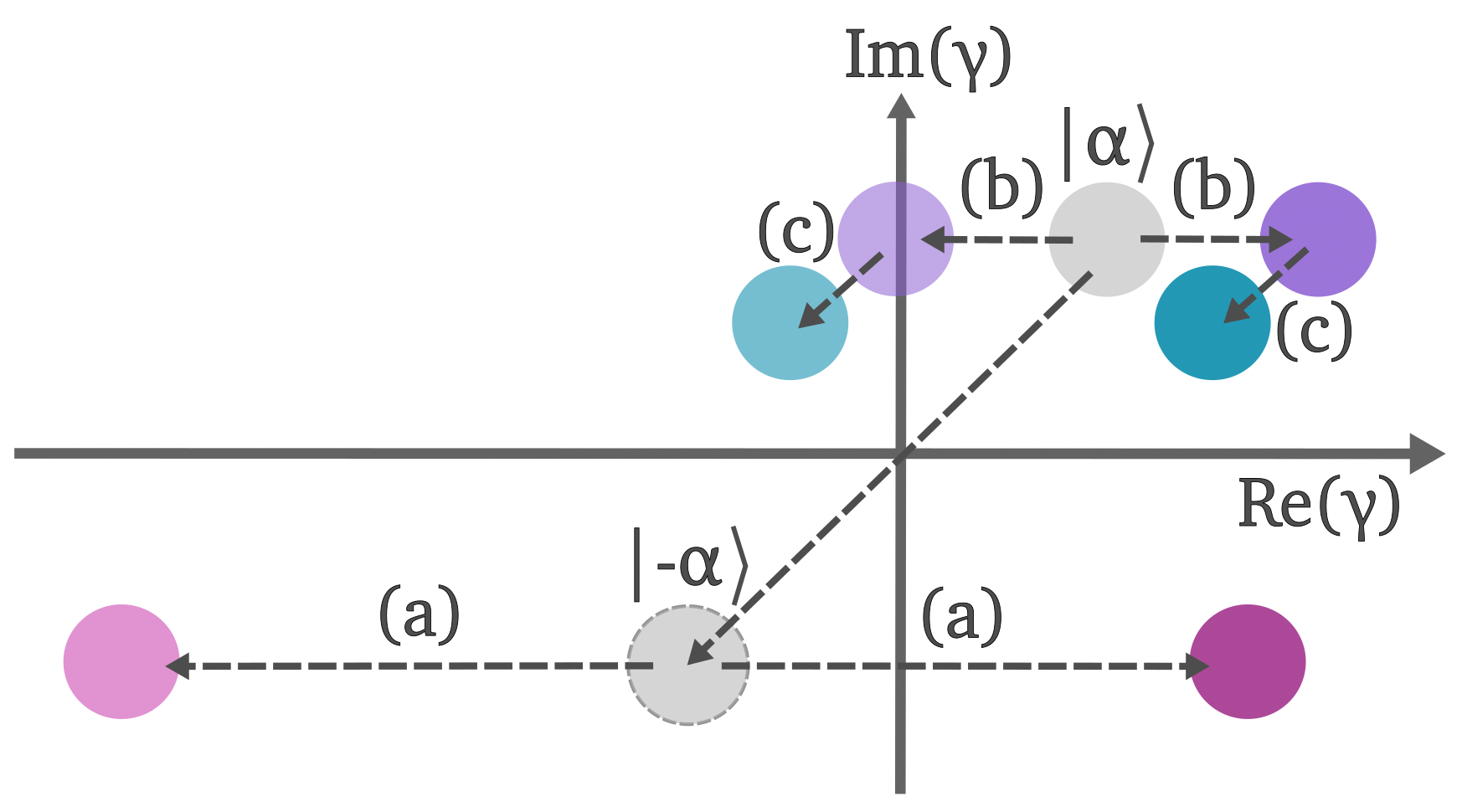}
    \caption{A visual representation of the resulting states that occur with pulse sequences, with and without a thermal bath. $\gamma$ refers to the parameters of an arbitrary coherent state $\ket{\gamma}$. Diagram is not to scale. \textbf{(a)} Depiction of Eq. \ref{eq:pulselessState}, where $\ket{-\alpha}$ (in grey) is translated by $\pm 2g/\omega$ (to pink). \textbf{(b)} Depiction of Eq. \ref{eq:pulseState}, where $\ket{\alpha}$ (in grey) is translated by $\pm (1/4)\omega g \tau^{2}$ (to purple). \textbf{(c)} Depiction of Eq. \ref{eq:pulseStateBath}, where the state in purple is further translated by $\theta_2$ (to blue; since we do not have an explicit form for $\theta_2$, this is an example depiction).}
    \label{fig:StatesWithPulses}
\end{figure}

Next, we must understand how pulse sequences affect the EW in Equation \ref{eq:NVwitness}. Interestingly, we can draw an equivalency between the state for the Ramsey sequence evaluated at $\tau = \pi/\omega$ and the Echo sequence evaluated at $\tau \ll \pi/\omega$. Thus, we attempt to modify the non-pulsed EW at $\tau = \pi/\omega$ such that it can be used when pulsed sequences are employed at $\tau \ll \pi/\omega$, which is reasonable given that determining the existence of entanglement within a short time duration is advantageous, especially given the short coherence times of NV spins. 

\begin{figure*}
\centering
    \sidesubfloat[]{
        \includegraphics[width=8.2 cm]{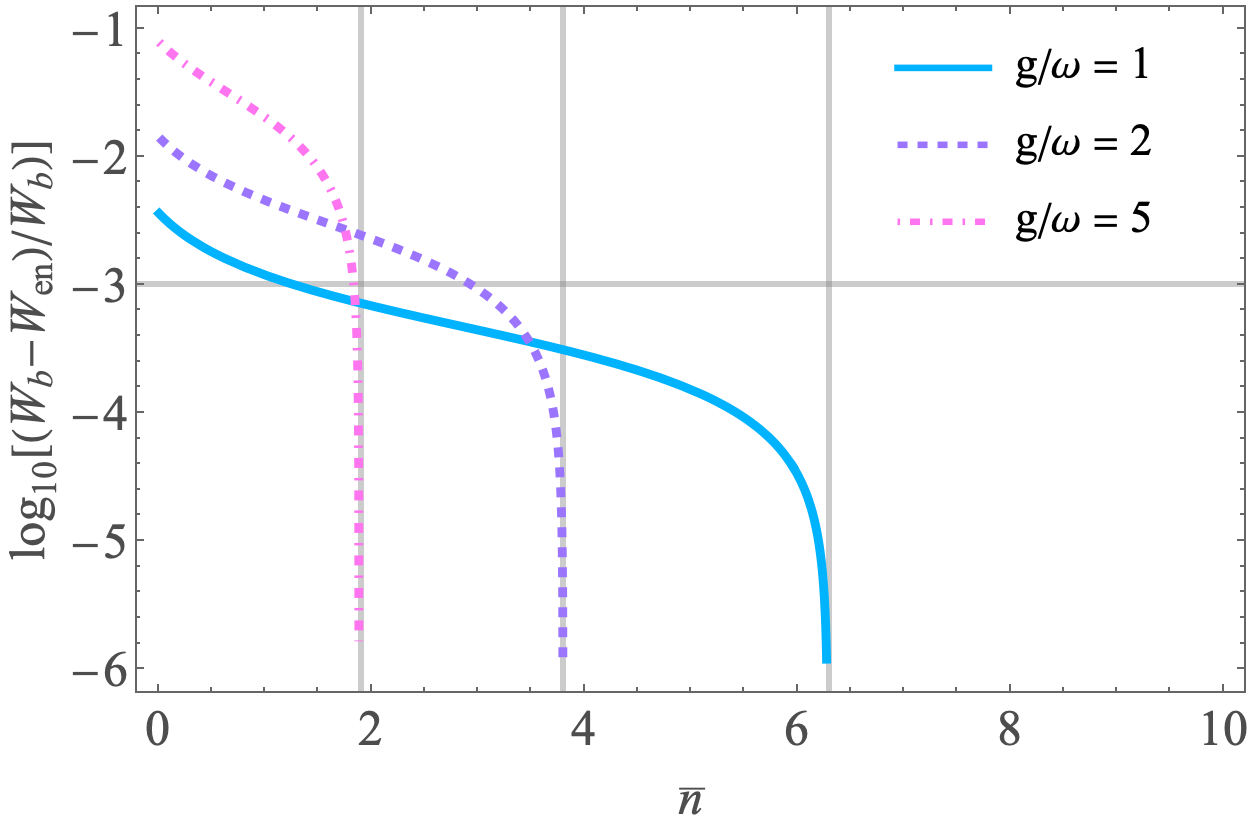}
        \label{fig:PulseEWInitThermState}
      }
    \sidesubfloat[]{
        \includegraphics[width=8.2 cm]{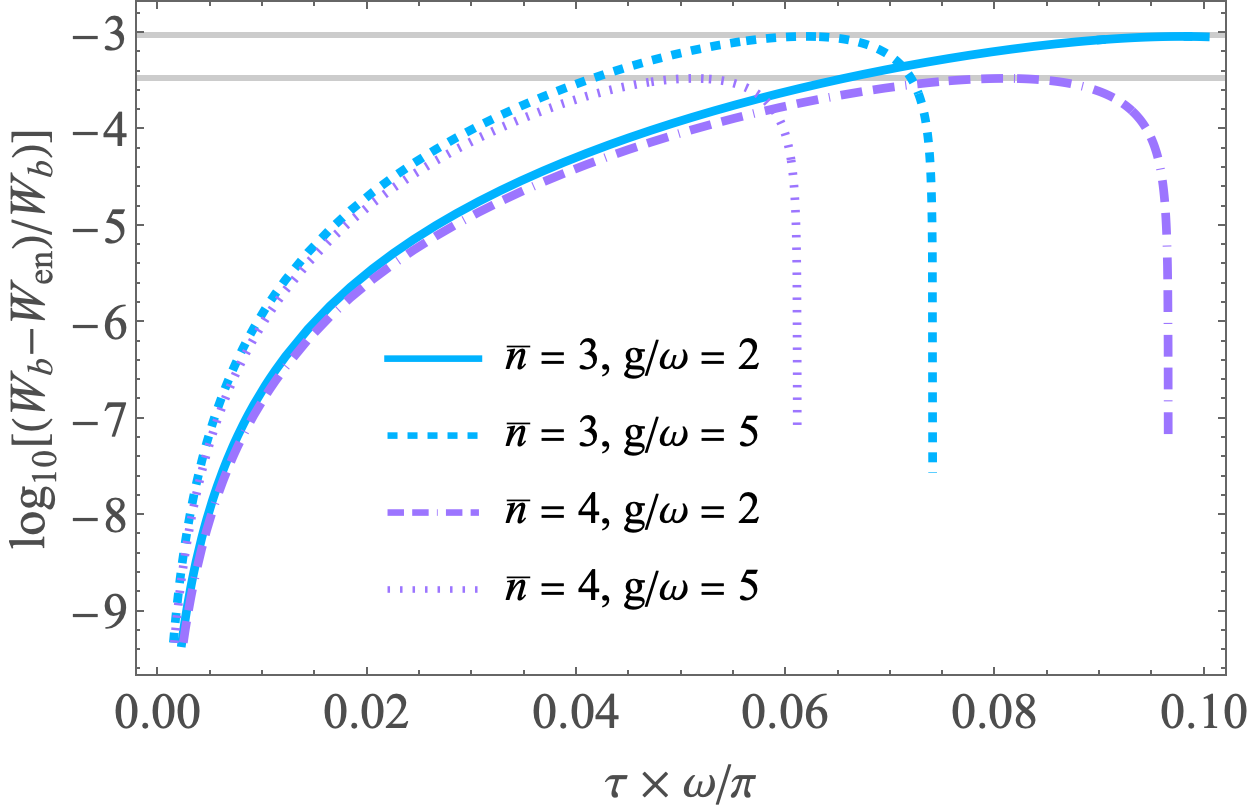}
        \label{fig:PulseViolationVsTau}
      }
     \hfill 
      \sidesubfloat[]{
        \includegraphics[width=8.2 cm]{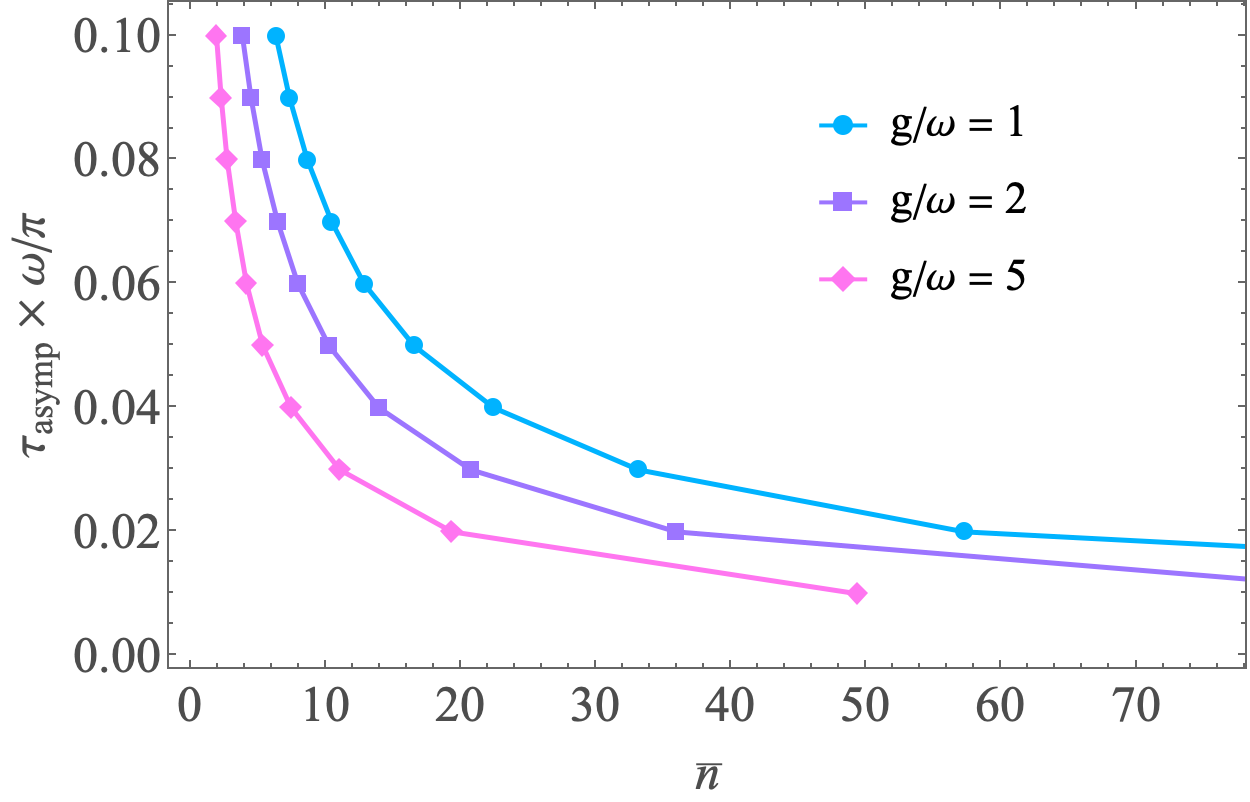}
        \label{fig:PulseEWInitThermStateTauVSnbar}
      }
      \sidesubfloat[]{
        \includegraphics[width=8.2 cm]{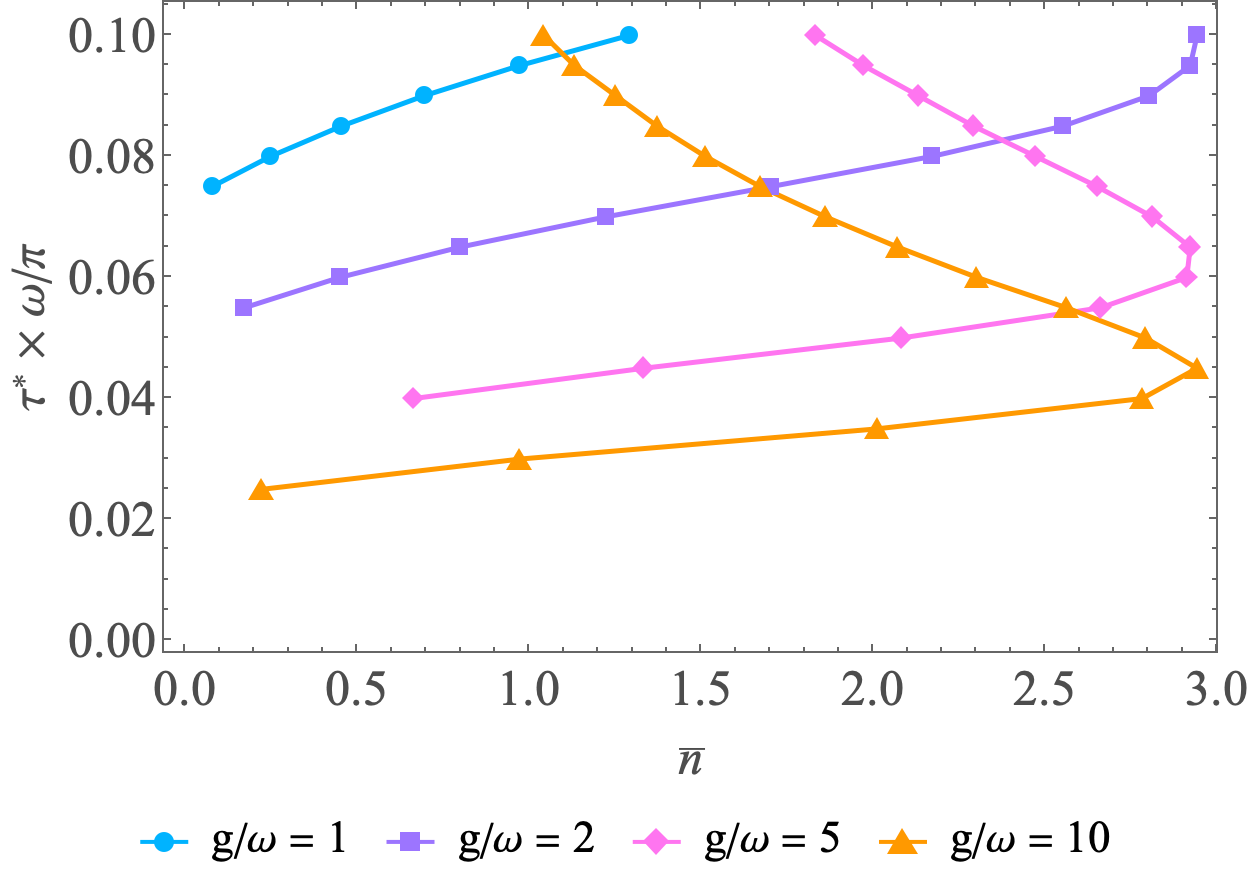}
        \label{fig:PulseEWInitThermStateTauStarVSnbar}
      }
    \caption{For all these figures, $\omega = 2 \pi \times 100$ Hz. \textbf{(a)} A plot of $\log_{10}[W_{\rm{ratio}}]$ against $\bar n$ for different values of $g$ (where $W_{\rm{ratio}} = (W_{b}-W_{\rm{en}})/W_{b}$). Here, $\lambda = (1/4)\omega g\tau^{2}$ and  $\tau = 0.1 \times \pi/\omega$. Note that the plots truncate (i.e. violation ceases to exist) at $\bar n \sim O(1)$. Moreover, the higher the value of $g/\omega$, the lower the value of $\bar n$ at which the violation stops. The vertical gray grid-lines emphasize the asymptotes, whereas the horizontal grid-line highlights a violation of $10^{-3}$ (which requires at least $10^{6}$ measurements.) \textbf{(b)} A plot of $\log_{10}[W_{\rm{ratio}}]$ against $\tau \times \omega/\pi$ for different values of $g/\omega$ at $\bar n = 3$ and $\bar n = 4$. Interestingly, the maximum violation appears to be the \textit{same} for a particular value of $\bar n$, regardless of the coupling. The higher the coupling, the sooner $W_{\rm{ratio}}$ reaches this maximum. \textbf{(c)} For different $g/\omega$, we extract $\tau \times \omega/\pi$ and $\bar n$ at the point where $\log_{10}[W_{\rm{ratio}}]$ asymptotes, i.e. violation of the EW ceases to exist. We label this time as $\tau_{\rm{asymp}}$. Thus, this plot tells us of the maximum time period for which violation exists as $\bar n$ increases. As expected, the higher the value of $\bar n$, the shorter the duration for which violation exists.  \textbf{(d)} For different $g/\omega$, we extract $\tau \times \omega/\pi$ and $\bar n$ at the point where $\log_{10}[W_{\rm{ratio}}]$ has a value of $-3$; this $\tau$ is represented as $\tau^{*}$. Hence, this figure shows us the time $\tau$ at which a violation $W_{\rm{ratio}}$ of \textit{at least} $10^{-3}$ is reached. As $\bar n$ decreases, there are two points where this occurs -- this is because the violation goes above $10^{-3}$ before decreasing eventually, thus resulting in two points at which $W_{\rm{ratio}}$ crosses $10^{-3}$. Interestingly, $\bar n = 3$ is the maximum value at which a violation of $10^{-3}$ can be reached for \textit{all} three of the highest couplings shown here. Both these points can also be seen in \textbf{(b)}. The data in sub-figs.~\textbf{(c)} and \textbf{(d)} have been extracted from plots similar to that of sub-figure \textbf{(a)}.}
    \label{fig:PulseEWInitThermStateAll}
\end{figure*}

If we consider the basic Hamiltonian without pulses to be 
\begin{equation}
    H_{0-1} =  g \sigma_z (a + a^\dag) + \omega a^\dag a,
\label{eq:NVhamilTwoLevelNoPulse}
\end{equation}
where we have dropped the Larmor precession and pulse terms of Eq. \ref{eq:NVhamilTwoLevel}, we find that the unitaries with and without pulses are given by 
{\small
\begin{align}
&U_{\rm{pulseless}}= D^{\dagger}(g \sigma_z/\omega)e^{-i \omega a^{\dagger}a\tau}D(g \sigma_z/\omega)
\label{eq:NoPulseUnitary}\\
&U_{\rm{pulse}}=D(g \sigma_z/\omega)e^{-i \omega a^{\dagger}a\tau/2} D(-2g \sigma_z/\omega)e^{-i \omega a^{\dagger}a\tau/2}D(g \sigma_z/\omega)
\label{eq:PulseUnitary}
\end{align}}
One can see that the spin-motion coupling term in the Hamiltonian appears in the displacement operators; thus, we expect that the coupling results in some displacement of the initial coherent state.

We can now apply these unitaries to the initial state $\ket{\alpha}(\ket{0}+\ket{1})/\sqrt{2}$, where $\ket{\alpha}$ is an arbitrary coherent state. At $\tau = \pi/\omega$ for $U_{\rm{pulseless}}$ and $\tau \ll \pi/\omega$ for $U_{\rm{pulse}}$, we get:
\begin{multline}
\ket{\psi}_{\rm{pulseless}} \approx\frac{1}{\sqrt{2}}\ket{-\alpha-\frac{2g}{\omega}}\ket{0}+\frac{1}{\sqrt{2}}\ket{-\alpha+\frac{2g}{\omega}}\ket{1}
\label{eq:pulselessState}
\end{multline}
and
\begin{multline}
\ket{\psi}_{\rm{pulse}} \approx \frac{1}{\sqrt{2}}\ket{\alpha \mathcal{H}(\omega,\tau)-\frac{1}{4} \omega g \tau^{2}}\ket{0}\\+\frac{1}{\sqrt{2}}\ket{\alpha\mathcal{H}(\omega,\tau)+\frac{1}{4} \omega g \tau^{2}}\ket{1}
\label{eq:pulseState}
\end{multline}
where $\mathcal{H}(\omega,\tau)$ is simply the expansion of $e^{-i \omega t}$ in $\omega \tau$; that is, $\mathcal{H}(\omega,\tau) =\left(1-\frac{1}{2} \omega^{2} \tau^{2}-i\omega\tau\right)$. Note that the phase terms which arise as a result of applying displacement operators sequentially have not been included;
furthermore, there is a phase difference between the $\alpha$ terms of  the two states (because the comparison is made at two different times) that we shall disregard. We think that these choices are acceptable for this particular discussion because the changes that concern us are the differences between the actual displacements of the states, rather than between their phases. In particular, the key difference between $\ket{\psi}_{\rm{pulseless}}$ and $\ket{\psi}_{\rm{pulse}}$ is the change $\pm 2g/\omega \rightarrow \pm (1/4)\omega g\tau^{2}$. This suggests that we can still use the results in Eqs. \ref{eq:ayhalfperiod}-\ref{eq:Wenhalfperiod}, but with $\lambda$ (where $\lambda = 2g/\omega$) now replaced with $(1/4)\omega g\tau^{2}$ (note that these equations only have terms of order $\lambda^{2}$; thus the sign of the displacement does not come into the final evaluation). Note that the previous EW results were for when the oscillator started in a thermal state, whereas the analysis in this section is for when it starts in an arbitrary coherent state; nevertheless, such a replacement of parameters follows through since a thermal state is composed of coherent states.

After modifying Eqs. \ref{eq:ayhalfperiod}-\ref{eq:Wenhalfperiod} by replacing $\lambda$ with $(1/4)\omega g\tau^{2}$, we plot the results in Figure \ref{fig:PulseEWInitThermStateAll}. Figure \ref{fig:PulseEWInitThermState} shows $\log_{10}[W_{\rm{ratio}}]$ vs $\bar n$ at $\tau = 0.1 \times \pi/\omega$ for different values of $g/\omega$. It can be seen that the plots asymptote as $\bar n$ increases, indicating the value of $\bar n$ at which the violation ceases to exist. Interestingly, the higher the coupling, the lower the value of $\bar n$ at which the asymptotes occur. This may appear counterintuitive, because it seemingly paints a higher coupling as being disadvantageous. 

Figure \ref{fig:PulseViolationVsTau} shows how the violation varies with $\tau$ for various combinations of $g/\omega$ and $\bar n$. Here we see how higher coupling provides an advantage -- a higher coupling allows the system to reach maximal violation sooner. Interestingly, this maximal violation appears to be independent of $g/\omega$, and is instead determined by $\bar n$. As expected, higher $\bar n$ results in a lower maximum. 

Figure \ref{fig:PulseEWInitThermStateTauVSnbar} plots the time at which the (log plot of the) violation asymptotes (i.e., violation ceases to occur) against $\bar n$. As expected, we see that increasing $\bar n$ reduces the duration for which violation exists; notably, the higher the value of $g/\omega$, the smaller the value of $\tau_{\rm{asymp}} \times \omega/\pi$ for a given value of $\bar n$. Next, in Fig. \ref{fig:PulseEWInitThermStateTauStarVSnbar}, we plot $\tau^{*} \times \omega/\pi$ vs $\bar n$, where $\tau^{*}$ is the time at which violation $(W_{b}-W_{\rm{en}})/W_{b}$ reaches $10^{-3}$ (that is, the time at which the required number of measurements to detect this violation reaches $10^{6}$). As $\bar n$ decreases, $W_{\rm{ratio}}$ crosses $10^{-3}$ at two points, because the maximum violation is above $10^{-3}$. Note that in these plots, $\tau \leq 0.1 \times \pi/\omega$, in order to meet the condition $\omega\tau \ll 1$, as per the expansion (and consequently, $\tau \ll \pi/\omega$).

That the maximum violation is independent of $g/\omega$ as shown by these graphs is an interesting feature. This behavior is consistent with the same effect seen in squeezed magnetometers and other quantum sensing systems: increasing coupling is also increasing sensitivity to noise, so the total performance vis entanglement is independent of $g/\omega$ (see Fig.~\ref{fig:PulseEWInitThermStateTauStarVSnbar}) \cite{PhysRevLett.104.013601,PhysRevLett.104.133601,PhysRevLett.92.230801}. The benefit instead is that higher $g$ leads to faster entanglement, which makes other effects such as spin dephasing less important. 

We now consider the effects of a thermal bath. As in Section \ref{SetupEW}, thermal noise is introduced as a force term, such that
\begin{equation}
    H_{\rm{bath}} =  \mathcal{G}(t)(a + a^\dag) + \omega a^\dag a.
\label{eq:NVhamilTwoLevelPulseBath}
\end{equation}
where
\begin{equation}
    \mathcal{G}(t) \equiv g(t) \sigma_z+f_{\rm{th}}(t).
\label{eq:CurlyG}
\end{equation}
Here $g(t) = g$ from $0-\tau/2$ and $g(t) = -g$ from $\tau/2-\tau$; the time-dependency reflects the use of pulse sequences, because $g$ essentially flips sign at $\tau/2$ ($\sigma_x\sigma_z\sigma_x=-\sigma_z$, thus effectively flipping the sign of $g$). 

For the rest of this section, we work within the interaction picture, for which the Hamiltonian is given by:
\begin{equation}
    \tilde{H}_{\rm{bath}}(t) = \mathcal{G}(t) (a e^{-i\omega t}+a^\dagger e^{i\omega t}).
\label{eq:HBath1}
\end{equation}
The time evolution operator can then be written as 
{\small
\begin{multline}
    \tilde{U}  = \exp[i\theta_{1,a}+i\theta_{1,b}]\exp\{-i\int_0^{\tau}dt'~\tilde{H}_{\rm{bath}}(t)\},
\label{eq:UnitaryAverage}
\end{multline}}
for which the phase factor is of the form
{\small
\begin{multline}
\exp[i\theta_{1,a}+i\theta_{1,b}]\\=\exp\{i \int_0^{\tau}\int_0^{t'}dt'dt''[g(t')g(t'')+ f_{\rm{th}}(t') f_{\rm{th}}(t'')]\sin[\omega(t'-t'')]\\
    + i \int_0^{\tau}\int_0^{t'}dt'dt'' \sigma_z[g(t')f_{\rm{th}}(t'')+g(t'')f_{\rm{th}}(t')]\sin[\omega(t'-t'')]\}.
\label{eq:UnitaryTheta}
\end{multline}}
where $\theta_{1,b}$ has the operator $\sigma_z$ in the integrand, indicating that there is a spin-dependent phase through which the noise $f_{\rm{th}}(t)$ enters.  Note that the first term, proportional to $g(t')g(t'')$, leads to squeezing for the case of many spins, and will discussed in the following section. 

Further simplification shows that the unitary can be rewritten as
\begin{multline}
    \tilde{U} = \exp[i\theta_{1,a}+i\theta_{1,b}]\\
    \times \exp\{[\sigma_z(\beta_1+\beta_2)+\theta_2]a^\dagger-[\sigma_z(\beta_1+\beta_2)^{*}+\theta_2^{*}]a\}\\
    = \exp[i\theta_{1,a}+i\theta_{1,b}]D[\sigma_z(\beta_1+\beta_2)+\theta_2]
\label{eq:UnitaryEval1}
\end{multline}
where
\begin{align}
    \beta_1 &= \frac{g}{\omega}(1-e^{i\omega \tau/2})\\
    \beta_2 &= \frac{g}{\omega}(e^{i\omega\tau}-e^{i\omega \tau/2})\\
    \theta_2 &= -i\int_0^{\tau}dt'~f_{\rm{th}}(t') e^{i\omega t'}\}.
\label{eq:Betas}
\end{align}
Hence, we see that the unitary evolution is composed of a displacement operator, along with exponentials that introduce phases to the states. $\theta_2$, which is a parameter of the displacement operator, is another method by which $f_{\rm{th}}(t)$ affects the state -- this time, the force causes a displacement of the state.

Next, we investigate the factor $\exp[i\theta_{1,a}+i\theta_{1,b}]$ and understand how the spin-dependent phase introduced by $\theta_{1,b}$ affects the final state. Taking the Fourier transform of the noise and evaluating the time integrals, $\theta_{1,b}$ can be written as 
\begin{equation}
    \theta_{1,b} = \int_{-\infty}^{\infty}\frac{d\nu}{\sqrt{2\pi}} \sigma_z f_{\rm{th}}(\nu) \mathcal{F}(\nu).
\label{eq:theta1bFourier}
\end{equation}

For a white noise force, where $\langle\langle f_{\rm{th}}(\nu)f_{\rm{th}}(\nu')\rangle\rangle = 2 \gamma \bar n \delta(\nu+\nu')$, the variance $\langle\langle \theta_{1,b}^{2} \rangle \rangle$ is given by 
\begin{equation}
    \langle\langle \theta_{1,b}^{2} \rangle \rangle = \int_{-\infty}^{\infty}\frac{d\nu}{2\pi} 2\gamma \bar n \mathcal{F}(\nu)\mathcal{F}(-\nu) = \int_{-\infty}^{\infty}\frac{d\nu}{2\pi} 2\gamma \bar n |\mathcal{F}(\nu)|^{2},
\label{eq:theta1bVariance}
\end{equation}
since $\mathcal{F}(\nu)\mathcal{F}(-\nu)=|\mathcal{F}(\nu)|^{2}$.  The exact expression for $\mathcal{F}(\nu)$ is given in Appendix \ref{EWwithPulsesApp}, along with a plot of $|\mathcal{F(\nu)}|^{2}$ for visualization in Figure \ref{fig:FilterFunction}.

Defining 
\begin{equation}
    \tilde{\theta}_{1,b} \equiv \int_{-\infty}^{\infty}\frac{d\nu}{\sqrt{2\pi}} f_{\rm{th}}(\nu) \mathcal{F}(\nu),
\label{eq:theta1bFourier}
\end{equation}
the final state, $\tilde{U}\ket{\alpha}(\ket{0}+\ket{1})/\sqrt{2}$, can be evaluated to give
{\small
\begin{multline}
\ket{\psi}_{\rm{pulse,bath,int}}
\approx\frac{\exp(i\theta_{1,a})}{\sqrt{2}}\\
\times [e^{i\tilde{\theta}_{1,b}}\ket{\alpha-\frac{1}{4}\omega g \tau^{2}+\theta_2}\ket{0}+e^{-i\tilde{\theta}_{1,b}}\ket{\alpha+\frac{1}{4}\omega g \tau^{2}+\theta_2}\ket{1}].
\label{eq:pulseStateBath}
\end{multline}}
Thus, we see that $f_{\rm{th}}(t)$ results in both an additional displacement of the initial coherent state and a spin-dependent phase. $\theta_{1,a}$ is simply a global phase. Note that Eq. \ref{eq:pulseStateBath} is in the interaction picture. We have not applied the approximation $\omega \tau \ll 1$ to the displacement caused by the force, since $f_{\rm{th}}(t)$ does not have an explicit form by itself and is instead characterized by its white noise auto-correlation function (at least, in our analysis). As previously, the phases arising from applying displacement operators have been dropped.
Further details of this calculation are given in Appendix \ref{EWwithPulsesApp}.

\section{Squeezing}
\label{Squeezing}

\begin{table*}[htb]
\caption{Squeezing for each pulse sequence relative to the spin-motion coupling $g$, taking $\xi = 1/4$. \label{t:squeeze}}
\begin{center}
\begin{tabular}{|l|c|c|}
\hline
Seq. & $\zeta/g^2$    & Squeezing at SQL \\
\hline 
Ramsey & 		$\frac{\omega \tau^3}{6}$		&  $\frac{\omega \tau}{6 N_s}$ \\
Hahn echo & 	$\frac{\omega \tau- 4 \sin(\omega \tau/2) + \sin(\omega \tau)}{\omega^2} \sim \frac{ \omega \tau^3}{12}$	 & 	$\frac{4}{3 \omega \tau N_s}$ \\
Carr-Purcell &	$\frac{\omega \tau - 4 \sin(\omega \tau/4) - 4 \sin(\omega \tau/2) + 4 \sin(3 \omega \tau/4) - \sin(\omega \tau)}{\omega^2} \sim \frac{\omega \tau^3}{48}$   & $\frac{64}{3 \omega^3 \tau^3 N_s}$  \\
\hline
\end{tabular}
\end{center}
\label{default}
\end{table*}%

The performance of complex pulse sequences only matters if we have sufficiently strong coupling to achieve the SQL, and if thermal noise and other sources are sufficiently weak. As shown above, this may not be relevant for the case of single spins. For example, a micrometer-sized diamond with 10 ppm NV centers has more than $N_s \sim 10^5$ NVs within it. Many of the results above immediately scale to the case of $N_s$ spins. Generically we expect for $n^*$ to increase by a factor of $N_s$ due to the independent sum of each spin's backaction, while $\Delta n$ is the same for a single spin's phase evolution. At the same time, the spin projection noise is reduced by $\sqrt{N_s}$. Thus the $N_S$-dependent noise to signal ratio, adapting from Eq.~\ref{eq:nsr}, is 
\begin{equation}
NSR_{\rm N} \sim \frac{N_s/4 + N_s (\Delta n)^2 \xi}{\phi^2}
\end{equation}
We see that this acts effectively as an increase of $g \rightarrow g \sqrt{N_s}$, as would be naively expected. The SQL is thus
\begin{equation}
g^* \sqrt{N_s} = \frac{1}{\sqrt{2 (\Delta n/g^2) \sqrt{\xi}}}
\end{equation}
For the case of Carr-Purcell, this yields $g^* \sqrt{N_s} \sim 2^{5}/\omega^2 \tau^3$.

However, when using $N_s$ spins, we also have a new term that emerges: the spin projection $J_z \equiv \sum_j S_z^j$ produces a force on the mechanical system, which in turn causes a rotation around the $z$ axis of the spins that is proportional to the spin projection. This is a type of squeezing, represented by $\exp(-i \zeta J_z^2)$ as a unitary operation. As shown in Appendix~\ref{a:magnus}, this $\zeta$ parameter is given by
\begin{equation}
\zeta = \int_0^\tau \int_0^t \sin(\omega (t-t')) g(t) g(t') dt' dt
\end{equation}
where $g(t)$ is the coupling that changes sign due to the various pulses used in a given sequence.

For our three sequences of choice, we have squeezing as shown in Table~\ref{t:squeeze}, with scaling $\zeta \sim g^2 \omega \tau^3$. We can now compare this with the force SQL, which gives a value of $g^*$, which is nominally ideal. We see that for $\omega \tau \ll 1$, squeezing at SQL from higher order pulse sequences is far larger than for smaller sequences. Even for modest $\omega \tau$, the large numerators and prefactors indicate that squeezing will be accessible in general and must be accounted for when doing sensing with many spins.

Now we can ask: can we use squeezing during the sequence to improve the force sensing capabilities of the system? We find the answer to be `yes' in practice, if we make a small rotation of the final observable -- rather than measuring, e.g. $J_y$, we instead do an additional rotation about the $x$ axis of the spin by an angle
\begin{equation}
\theta = -\tan^{-1}\left( \frac{4}{N_s \zeta} + \frac{N_s \zeta}{2} \right)
\end{equation}
This improves the signal to noise by reducing the spin shot noise as
\begin{equation}
\frac{1}{\sqrt{N_s}} \rightarrow \frac{1}{\sqrt{N_s} \sqrt{1 + (N_s \zeta/4)^2}}
\end{equation}
for $\zeta N_s \ll \sqrt{N_s}$. The limit on large $\zeta N_s$ reflects a breakdown of the bosonic approximation to the spin variables $J_y$ and $J_z$. This also confirms that this approach is unlikely to approach the Heisenberg limit, though it is a practical improvement.

Thus, our system combines both backaction evasion and sub-Poissonian statistics, which occurs naturally in the high Q limit so long as the phonon heating rate is sufficiently low, as may be realized in levitated systems. We note that these benefits are also limited by, e.g., $T_2$ for Carr-Purcell sequences and will depend sensitively on many specific parameters of the system.

\section{Summary}

In this paper, we proposed a sensing system composed of an NV center spin coupled to the motion of its levitated host diamond, where the interaction between the spin and the diamond results from the gradient of the magnetic trap. We find that shot noise limits force sensitivity more than room temperature thermal noise and that backaction is significant enough to be addressed. Backaction evasion with a Carr-Purcell-type sequence can significantly enhance the sensitivity of this sensor to forces; in particular, we find that we can obtain sensitivities $< 10^{-23}$ N/$\sqrt{\rm Hz}$ at frequencies of $10^{4}$ Hz.  We also investigate the generation of entanglement between the spin and the motion of the diamond using an entanglement witness (EW); once appropriate modifications are made, the same witness that is used for the case without pulses can be used for the Echo sequence. As expected, when starting in a thermal state, increasing the thermal occupancy ($\bar n$) results in a shorter duration for which the EW is violated. Thus, if entanglement generation is a priority, it is advantageous to start with a diamond that has been cooled to the ground state. Increasing the coupling results in the maximum violation being achieved faster, thus highlighting the importance of high coupling. 

There are unanswered theoretical questions, most crucially an EW analysis for the multi-spin case. This analysis will have to be conducted to all orders of $\lambda$ because $\lambda\gtrsim 1$, which is much more complicated mathematically. Furthermore, we have only expanded the EW analysis to incorporate the Echo sequence, even though in the future, it may be necessary to consider how it is affected by Carr-Purcell sequences as well.

A setup such as this has practical applications in force sensing and testing the quantum nature of forces. Indeed, our experimental-theoretical collaboration is attempting to practically implement a force sensor as described here. We would also like to determine if our apparatus can result in spin-motion entanglement, although we may have to use an alternative method to verify the generation of entanglement, given the practical challenges of measuring both the position and momentum of the diamond for the current witness in a time fast enough compared to the mechanical dephasing time. Furthermore, based on our analysis, it is necessary to achieve ground state cooling for accessible entanglement.

\section{Acknowledgments}

This work was supported by the John Templeton Foundation (Grant No. 63121). G.P. also acknowledges support from the Heising-Simons Foundation (Grant No. 2023-4467), while G.D., J.B., and D.P. were further supported by the Sloan Foundation (Grant No. G-2023-21131).

\appendix

\section{Pulsed evolution picture}
\label{a:magnus}

Consider the evolution under a sequence of $n$ microwave pulses $P_i$ applied to the spins. We write the total evolution as a series of unitary evolutions interleaved with pulses, as
\begin{equation}
U(t_f,t_0) = U(t_f,t_n) P_n U(t_n,t_{n-1}) P_{n-1} \ldots P_1 U(t_1,t_0)
\end{equation}
From this we can define an interaction picture between $P_i$ and $P_{i+1}$ as 
\[
U_I(t_{i+1},t_i) = (\prod_{j \leq i} P_j)^\dag U(t_{i+1},t_i) (\prod_{j \leq i} P_j)
\]
In the case of only $\pi$ pulses, this corresponds to changes to sign of the pseudo spin $\sigma_z = \ket{0}\bra{0} - \ket{1}\bra{1}$. Thus we can write the Hamiltonian $H_I(t)$ in this interaction picture:
\begin{equation}
H_I = [-f(t) + g(t) \sigma_z] (a e^{-i \omega t} + a^\dag e^{i \omega t})
\end{equation}
where the pulses determine the sign of $g$ over time.

\begin{figure}
    \centering
    \includegraphics[width=8.6 cm]{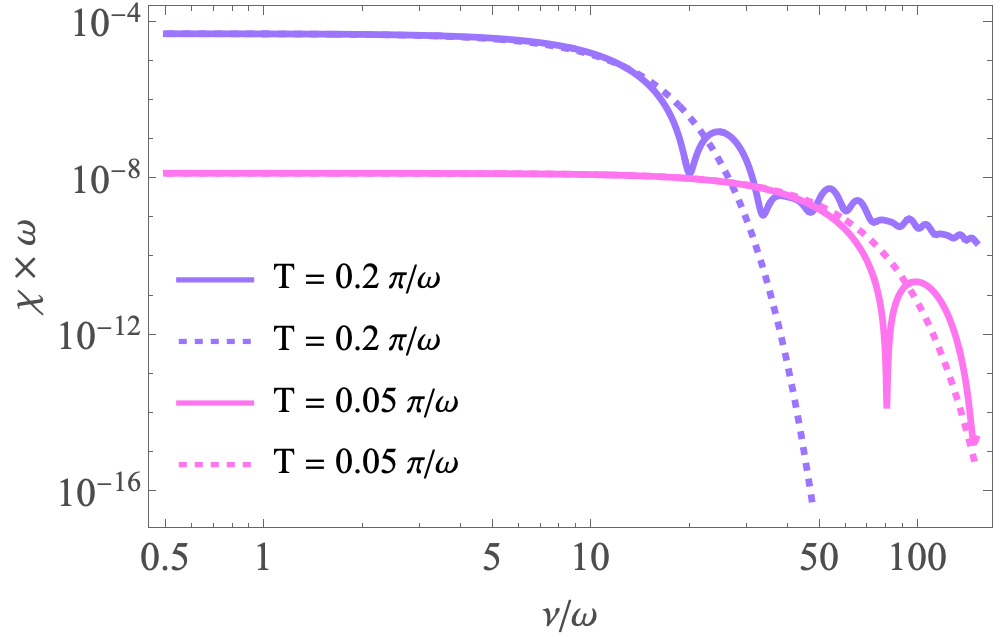}
    \caption{Numerical evaluation of the approximation (Eqn.~\ref{eq:PhiCP} appropriate for the response function $\chi(\nu)$ of the Carr-Purcell sequence (of length $T/4$) for $T = 0.2~ \pi/\omega$ (purple) and $T=0.05 ~\pi/\omega$ (pink). The plot is made unitless by comparing $\nu$ to $\omega$ and $\chi$ to $\omega^{-1}$. The approximation is appropriate for frequencies comparable to or below the bandpass $2 \pi/T$.}
    \label{fig:SpectralFunctionPlot}
\end{figure}

Within this formalism, we note that we can find the exact form of $U$ by using the Magnus expansion, as $[H_i(t),[H_i(t'),H_i(t'')]] = 0$. That is, 
\begin{align}
U(t_{f},t_0) &= \exp[-i (\Phi_1 + \Phi_2)] \\
\Phi_1 &= \int_{t_0}^{t_{f}} H_I(t') dt' \\
\Phi_2 &= \frac{1}{2 i} \int_{t_o}^{t_{f}} \int_{t_o}^{t'} [H_I(t'),H_I(t'')] dt'' dt'
\end{align}
Exploring these two terms, we have
\begin{equation}
\Phi_1 = \int_{t_0}^{t_{f}} (-f(t') + g(t') \sigma_z) e^{-i \omega t'} dt' a + {\rm H.c.}
\end{equation}
and thus this term contributes an overall (spin-dependent) displacement to the harmonic oscillator from the spin and from the external force. 

We now turn to the second term, which combines the effects of the force and of the spin-dependent interaction -- this yields a rotation of the spins that depends upon the force, and thus is our force signal of interest. 
Specifically, 
\begin{multline}
\Phi_2 = \int_{t_o}^{t_{f}} \int_{t_o}^{t'} \sin(\omega (t'-t'')) (-f(t') + g(t') \sigma_z)\\ \times (-f(t'') + g(t'') \sigma_z) dt'' dt'\ .
\end{multline}
If we look at an ensemble of spins with $J_z = 1/2 \sum_i \sigma_z^i$, then we see that $\Phi_2$ has two main contributions: phase evolution from the $f g$ cross term, and squeezing from the $g(t') g(t'')$ term. The term proportional to $f(t') f(t'')$ is a global phase and not observable.
{\small
\begin{multline}
\Phi_2 = -2 g_i J_z \int_{t_i}^{t_{i+1}} \int_{t_i}^{t'} \sin(\omega (t'-t'')) (f(t') + f(t''))dt'' dt' \\ +
4 g_i^2 J_z^2 \frac{\tau_i \omega - \sin(\tau_i \omega)}{\omega^2}
\label{eq:Phi2}
\end{multline}}
where $\tau_i = t_{i+1} - t_i$. This second term has a phase evolution under $J_z$ that is determined by the external force, and a squeezing term $\propto J_z^2$.

We remark that the approach we use in this paper -- which neglects explicit mechanical damping -- is appropriate when we are in the high $Q$ and high temperature regime $k_b T \gg \hbar \omega$ for the mechanical system. In this limit, the dominant effect of the mechanical environment is that of classical Brownian motion noise. That is,
\begin{equation}
f(t) \rightarrow f(t) + f_{\rm th}(t)
\end{equation}
with
\begin{equation}
\langle\langle{f_{\rm th}(t) f_{\rm th}(t')}\rangle\rangle = 2 \gamma \frac{k_b T}{\hbar \omega} \delta(t-t')
\end{equation}
In SI units, this is the usual $4 m \gamma k_b T \delta(t-t')$. 

This high temperature approximation captures the dominant errors in measurement.

\section{Entanglement witness elaboration}
\label{EWAppendix}

In this section, we provide more details on the entanglement witness (EW) calculation. We adopt the calculations in Ref. \cite{PremawardhanaEW}, after incorporating suitable modifications. Thus, some of the expressions presented in Sections \ref{NoiselessEWapp} and \ref{ThermalBathApp} are taken nearly verbatim from Ref. \cite{PremawardhanaEW} -- we include these expressions to make this paper self-contained. However, our results are now applicable to only one spin and all values of $\lambda$, as opposed to $N$ spins and up to $\lambda^{2}$ in Ref. \cite{PremawardhanaEW}. 

In Section \ref{SetupEW}, we gave expressions for quantities such as $W_b$ and $W_{\rm{en}}$. Since the \textit{general} expression for $W_b$ is a derivation taken from Ref. \cite{PremawardhanaEW} and does not depend on system parameters, we do not detail it here. Instead, we discuss the calculation of $W_{\rm{en}}$ under the various noise scenarios and how $a_{\mu}$ and $b_{\mu}$ were obtained.

When evaluating $W_{\textrm{en}}$ under various noisy situations, it is necessary to determine terms such as $\bra{\psi_{\textrm{in}}} \sigma_x(t)^{2} \ket{\psi_{\textrm{in}}}$, $\bra{\psi_{\textrm{in}}} \sigma_y(t) q(t) \ket{\psi_{\textrm{in}}}$, $\bra{\psi_{\textrm{in}}}\sigma_y(t) p(t) \ket{\psi_{\textrm{in}}}$ etc, with $\ket{\psi_{\textrm{in}}} = \ket{\psi_{s,\textrm{in}}} \otimes \ket{\psi_{\textrm{osc},\textrm{in}}}$, where $\ket{\psi_{s,\textrm{in}}}$ is the initial state of the spin and $\ket{\psi_{\textrm{osc},\textrm{in}}}$ is the initial state of the oscillator (diamond). For this, it is convenient to consider the Heisenberg evolution of operators under the appropriate Hamiltonian, as we shall do in the upcoming sections.

\subsection{Entanglement witness for initial thermal state}
\label{NoiselessEWapp}

Here, we expand on the evaluation of the noiseless EW. Defining $\tilde S_z = \sigma_z/2$, the spin operators evolve as follows \cite{PremawardhanaEW}:
\begin{align}
\tilde S_x(t) = \tilde S_x \cos(\phi_\textrm{ent})-\tilde S_y\sin(\phi_\textrm{ent})
\label{eq:SxHeis}\\
\tilde S_y(t) = \tilde S_y \cos(\phi_\textrm{ent})+\tilde S_x\sin(\phi_\textrm{ent}),
\label{eq:SyHeis}
\end{align}
where 
\begin{align}
\phi_{\rm ent} & = \omega_L t +\int_0^t 2g [a(t')+a^\dagger(t')] dt'
\label{eq:PhiEnt}\\
a(t) &= a_0 e^{-i \omega t} - i \int_0^t e^{-i \omega(t - t')} 2g \tilde S_z ~dt'. 
\end{align}
This gives \cite{PremawardhanaEW}
\begin{equation}
\phi_\textrm{ent} \approx \omega_L t+ \sqrt{\frac{2}{m \omega}}~\lambda p [1-\cos(\omega t)]+\sqrt{2 m \omega}~\lambda q \sin(\omega t),
\label{eq:Phi}
\end{equation}
where $\lambda=2 g/\omega$. Note that here, as in Ref. \cite{PremawardhanaEW}, we have dropped a term linear in $\tilde{S}_{z}$. Next, the expressions for the oscillator operators are \cite{PremawardhanaEW}:
\begin{align}
q(t) &= q \cos[\omega t] + \frac{p}{m \omega} \sin[\omega t]+\sqrt{\frac{2}{m \omega}} \lambda \tilde S_z [\cos(\omega t)-1]
\label{eq:qHeis}\\
p(t) &= p \cos(\omega t) - m \omega q \sin(\omega t)-\sqrt{2 m \omega} \lambda \tilde S_z \sin(\omega t).
\label{eq:pHeis}
\end{align}
Using Eqs. \ref{eq:SxHeis}-\ref{eq:pHeis} and expanding, we are left to evaluate terms such as $\bra{\psi_{\textrm{in}}}O_{s}O_{\textrm{osc}} \ket{\psi_{\textrm{in}}}$, which, as a result of $\ket{\psi_{\textrm{in}}}$ being separable, ultimately requires us to determine $\bra{\psi_{\textrm{s,in}}}O_{s} \ket{\psi_{\textrm{s,in}}}$ and $\bra{\psi_{\textrm{osc,in}}}O_{\textrm{osc}} \ket{\psi_{\textrm{osc,in}}}$.

Once $W_{\textrm{en}}$ is in hand, we consider the quantity \cite{PremawardhanaEW} 
\begin{equation}
W_{\textrm{ratio}} = \frac{W_{b}-W_{\textrm{en}}}{W_{b}}
\label{eq:Wratio}
\end{equation}
since the number of measurements required to determine this violation is $\sim W_{\textrm{ratio}}^{-2}$. We choose the coefficients $a_{\mu}$ and $b_{\mu}$ as explained in Ref. \cite{PremawardhanaEW}; we choose by solving for $a_{\mu}$ and $b_{\mu}$, such that $\partial_{a_{\mu}}W_{\textrm{en}}=0$ and $\partial_{b_{\mu}}W_{\textrm{en}}=0$. Thus, we obtain the expressions for the single-spin case where the oscillator starts in a thermal state:
{\small
\begin{equation}
W_{b}=\frac{1}{2}+\frac{e^{-(2\bar n +1)\lambda^{2}[1-\cos(\omega t)]}\cos (\omega_L t)\lambda^{2}[1-\cos(\omega t)]}{2\bar n +1+2 \lambda^{2}[1-\cos(\omega t)]}
\label{eq:WbInitTherm}
\end{equation}
\begin{multline}
    W_{\rm{en}}= \frac{1}{2}+\frac{1+2\bar n}{4(1+2 \bar n+2 \lambda^{2}-2 \lambda^{2}\cos(\omega t))}\\-\frac{e^{-2(1+2\bar n)\lambda^{2}(1-\cos(\omega t))}}{4}[1+2(1+2\bar n)\lambda^{2}-2(1+2\bar n)\lambda^{2}\cos(\omega t)]
\end{multline}}
These expressions hold for all values of $\lambda$ and do not contain the effects of the dropped $\tilde{S}_{z}$ term in $\phi_{\rm{ent}}$.

\subsection{Entanglement witness with white noise bath}
\label{ThermalBathApp}

For the scenario where the oscillator starts in a thermal bath, we must recalculate $W_{\rm{en}}$ and $W_{b}$.  

With \cite{PremawardhanaEW}
\begin{align}
\phi_{\rm ent,th} &= \omega_L t +\int_0^t 2g [a(t')+a^\dagger(t')]~ dt'  \\
a(t) &= a_0 e^{-i \omega t} - i \int_0^t e^{-i \omega(t - t')} (2g \tilde S_z+f_{\textrm{th}}(t')) ~dt',
\end{align}
the noise contribution to the position, momentum, and rotation angle will then be \cite{PremawardhanaEW}:
\begin{align}
\mathcal{Q} & = - 2 q_{\textrm{zpf}} \int_0^t f_{\textrm{th}}(t') \sin[\omega(t-t')]~dt',\\
\mathcal{P} & = - 2 p_{\textrm{zpf}} \int_0^t f_{\textrm{th}}(t') \cos[\omega(t-t')]~dt',
\end{align}
and
\begin{multline}
    \Phi = 2g \int_0^t \frac{\mathcal{Q}(t')}{q_{\textrm{zpf}}}~dt'\\= -4g \int_0^{t}\int_0^{t'} f_{\textrm{th}}(t'') \sin(\omega(t'-t''))~dt'' dt'.
\end{multline}
With these, and using \cite{PremawardhanaEW},
\begin{equation}
\langle\langle{f_{\rm th}(t) f_{\rm th}(t')}\rangle\rangle = 2 \gamma \bar n \delta(t-t'),
\end{equation}
where $\bar n = k_b T/\hbar\omega$,
we find that the total noise contribution is \cite{PremawardhanaEW}:
\begin{align}
\Delta\langle\langle{\textrm{Var}(\tilde S_{x})}\rangle\rangle& =\frac{1}{2}\lambda^{2}\frac{\bar n}{Q}[6 \omega t - 8 \sin(\omega t)+\sin(2 \omega t)]\\
\Delta\langle\langle q^{2}\rangle\rangle & = 2 q_{\textrm{zpf}}^{2}\frac{\bar n}{Q}[2 \omega t - \sin(2 \omega t)]\\
\Delta \langle\langle p^{2} \rangle\rangle & = 2 p_{\textrm{zpf}}^{2}\frac{\bar n}{Q}[2 \omega t + \sin(2 \omega t)]\\
\Delta \langle\langle qp+pq \rangle\rangle & =4\frac{\bar n}{Q}\sin^{2}(\omega t)\\
\Delta \langle\langle \tilde S_{y}q+q \tilde S_{y} \rangle\rangle & =-16 \lambda\frac{\bar n}{Q}q_{\textrm{zpf}}\sin^{4}(\omega t/2)\\
\Delta \langle\langle \tilde S_{y}p+p \tilde S_{y} \rangle\rangle & = 8 \lambda\frac{\bar n}{Q}p_{\textrm{zpf}}\left[\frac{\omega t}{2}-\sin(\omega t)+\frac{\sin(2 \omega t)}{4}\right]
\end{align}
We add these changes as described in Ref. \cite{PremawardhanaEW} in order to obtain the violation plots of Figure \ref{fig:EWbathPlots}:
{\small
\begin{multline}
    \Delta W_{\rm{bath}}=\Delta\langle\langle{\textrm{Var}(\tilde S_{x})}\rangle\rangle+a_{y}\Delta \langle\langle \tilde S_{y}q+q \tilde S_{y} \rangle\rangle+b_{y}\Delta \langle\langle \tilde S_{y}p+p \tilde S_{y} \rangle\rangle\\+(a_{y}^{2}+a_{z}^{2})\Delta\langle\langle q^{2}\rangle\rangle+(b_{y}^{2}+b_{z}^{2})\Delta \langle\langle p^{2}\rangle\rangle+(a_y b_y+a_z b_z)\Delta \langle\langle qp+pq \rangle\rangle
\label{eq:ThermBathWitChangeApp}
\end{multline}}

\subsection{Entanglement witness with pulse sequences}
\label{EWwithPulsesApp}

In this section, we elaborate on the steps necessary to determine how the EW must be modified to take into account the use of pulse sequences.

The Hamiltonian 
\begin{equation}
    H_{0-1} =  g \sigma_z (a + a^\dag) + \omega a^\dag a,
\label{eq:NVhamilTwoLevelNoPulseApp}
\end{equation}
can be rewritten as
\begin{equation}
    \tilde{H}_{0-1} = \omega(a^{\dagger}+g\sigma_{z}/\omega)(a+g\sigma_{z}/\omega)-g^{2}\sigma_z^{2}/\omega.
\label{eq:NVhamilTwoLevelNoPulseApp}
\end{equation}
With this, we can evaluate the unitary 
\begin{equation}
U_{\rm{pulseless}}= e^{-iH\tau}
\label{eq:NoPulseUnitaryApp0}
\end{equation}
to obtain
\begin{equation}
U_{\rm{pulseless}}= D^{\dagger}(g \sigma_z/\omega)e^{-i \omega a^{\dagger}a\tau}D(g \sigma_z/\omega).
\label{eq:NoPulseUnitaryApp}
\end{equation}
The sequence 
\begin{align}
U_{\rm{pulse}}&= \sigma_x U_{\rm{pulseless}}(\tau/2,\tau)\sigma_x U_{\rm{pulseless}}(0,\tau/2)
\end{align}
is used to get the unitary in Eq. \ref{eq:PulseUnitary},
{\small
\begin{align}
U_{\rm{pulse}}&=D(g \sigma_z/\omega)e^{-i \omega a^{\dagger}a\tau/2} D(-2g \sigma_z/\omega)e^{-i \omega a^{\dagger}a\tau/2}D(g \sigma_z/\omega).
\label{eq:PulseUnitaryApp}
\end{align}}

The complete expressions for the states after applying the unitaries in Eq. \ref{eq:NoPulseUnitaryApp} and \ref{eq:PulseUnitaryApp} are
\begin{multline}
\ket{\psi}_{\rm{pulseless}}=\frac{1}{\sqrt{2}}\ket{(\alpha+g/\omega)e^{-i\omega \tau}-g/\omega}\ket{0}\\+\frac{1}{\sqrt{2}}\ket{(\alpha-g/\omega)e^{-i\omega \tau}+g/\omega}\ket{1}
\label{eq:pulselessStateApp}
\end{multline}
\begin{multline}
\ket{\psi}_{\rm{pulse}}=\frac{1}{\sqrt{2}}\ket{(\alpha+g/\omega)e^{-i\omega \tau}+g/\omega-\frac{2g}{\omega}e^{-i\omega \tau/2}}\ket{0}\\+\frac{1}{\sqrt{2}}\ket{(\alpha-g/\omega)e^{-i\omega \tau}-g/\omega+\frac{2g}{\omega}e^{-i\omega \tau/2}}\ket{1}.
\label{eq:pulseStateApp}
\end{multline}
These expressions can be expanded for $\omega \tau \ll 1$ to obtain Equations \ref{eq:pulselessState} and \ref{eq:pulseState}.

Next we consider the effects of a thermal bath. The Hamiltonian
\begin{equation}
    H_{\rm{bath}} =  [g(t) \sigma_z+f_{\rm{th}}(t)] (a + a^\dag) + \omega a^\dag a.
\label{eq:NVhamilTwoLevelPulseBath}
\end{equation}
can be transformed into the interaction picture using \cite{AxlineThesis}
\begin{equation}
    \tilde{H}_{\rm{bath}} = i\dot{\mathcal{U}}\mathcal{U}^\dagger+\mathcal{U}H\mathcal{U}^{\dagger},
\label{eq:InterPicTransform}
\end{equation}
where $\mathcal{U}(t)=e^{i\omega a^\dagger at}$. Then, the interaction picture Hamiltonian is
\begin{equation}
    \tilde{H}_{\rm{bath}}(t) = \mathcal{G}(t) (a e^{-i\omega t}+a^\dagger e^{i\omega t})
\label{eq:HBath1}
\end{equation}
where
\begin{equation}
    \mathcal{G}(t) \equiv g(t) \sigma_z+f_{\rm{th}}(t).
\label{eq:CurlyG}
\end{equation}

\begin{figure}
    \centering
    \includegraphics[width=8.6 cm]{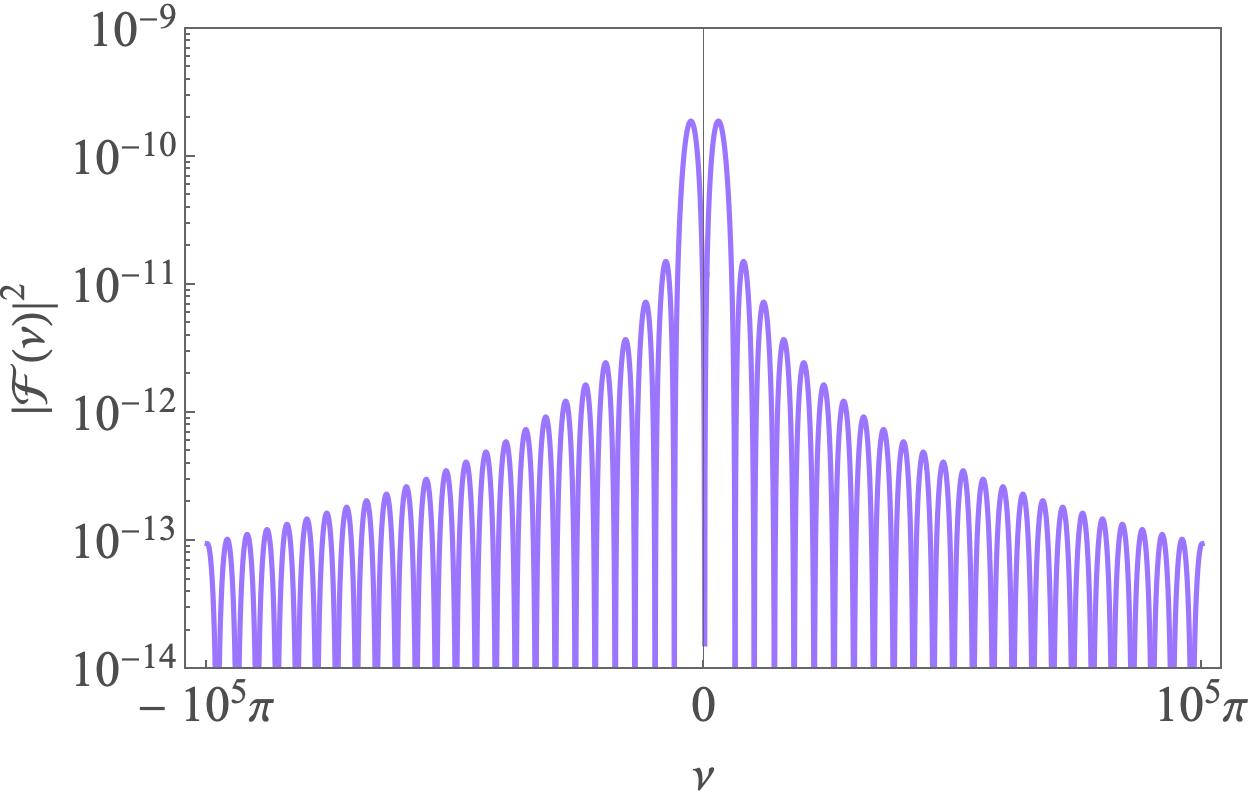}
    \caption{Visualization of $|\mathcal{F}(\nu)|^{2}$ for $g/\omega = 2$, $\omega = 2 \pi \times 100$ Hz, and $\tau = 0.1 \times \pi/\omega$. Note that the function is 0 at $\nu = 0$.}
    \label{fig:FilterFunction}
\end{figure}
The time evolution unitary in the interaction picture is then \cite{MagnusExpansion}
{\small
\begin{equation}
    \tilde{U}  = \mathcal{T}\left\{\exp\left[-i\int_0^{\tau}dt'~\tilde{H}_{\rm{bath}}(t')\right] \right\}.
\label{eq:InterPicHamil}
\end{equation}}
As we showed in Appendix, A, we can then write this unitary as 
\begin{multline}
    \tilde{U}  = \exp\{-i\int_0^{\tau}dt'~\tilde{H}_{\rm{bath}}(t')\\
    ~~~~~~~~~~+i\int_0^{\tau}\int_0^{t'}dt'dt''\mathcal{G}(t')\mathcal{G}(t'')\sin[\omega(t'-t'')]\}.
\label{eq:MagnusExp}
\end{multline}
This is exact because the second term only has the operator $\sigma_z$ and so the subsequent terms of the expansion are zero. This also means that it commutates through all the terms in the unitary, allowing it to be written as a separate exponential,
{\small
\begin{multline}
\exp[i\theta_{1,a}+i\theta_{1,b}]\\=\exp\{i \int_0^{\tau}\int_0^{t'}dt'dt''[g(t')g(t'')+ f_{\rm{th}}(t') f_{\rm{th}}(t'')]\sin[\omega(t'-t'')]\\
    + i \int_0^{\tau}\int_0^{t'}dt'dt'' \sigma_z[g(t')f_{\rm{th}}(t'')+g(t'')f_{\rm{th}}(t')]\sin[\omega(t'-t'')]\}.
\label{eq:UnitaryThetaApp}
\end{multline}}

Taking the Fourier transform of $f(t) = \int_{-\infty}^{\infty} f(\nu) e^{-i \nu t} \frac{d\nu}{\sqrt{2 \pi}}$, we have
\begin{multline}
    \theta_{1,b} = \int_{-\infty}^{\infty}\frac{d\nu}{\sqrt{2\pi}} \sigma_z f_{\rm{th}}(\nu)\\\int_0^{\tau}\int_0^{t'}dt'dt''
    [g(t')e^{-i\nu t''}+g(t'')e^{-i\nu t'}]\sin[\omega(t'-t'')]
\label{eq:theta1bFourierApp}
\end{multline}
which can then be simplified to obtain
\begin{equation}
    \theta_{1,b} = \int_{-\infty}^{\infty}\frac{d\nu}{\sqrt{2\pi}} \sigma_z f_{\rm{th}}(\nu) \mathcal{F}(\nu).
\label{eq:theta1bFourierApp2}
\end{equation}
The exact expression of $\mathcal{F}(\nu)$ is
\begin{multline}
    \mathcal{F(\nu)}=-\frac{2ie^{-\frac{1}{2}i\nu \tau}g}{\nu\omega(\nu-\omega)(\nu+\omega)}\\
    \times \{\cos(\nu\tau/2)[\nu^{2}-2\omega^{2}+\nu^{2}(-2\cos(\omega \tau/2)+\cos(\omega \tau))]\\
    +2\omega[\omega-4\nu\cos(\omega \tau/4)]\sin(\nu \tau/2)\sin^{3}(\omega \tau/4)\}
\label{eq:InterPicHamil}
\end{multline}
and the plot of $|\mathcal{F}(\nu)|^{2}$ against $\nu$ is given in Figure \ref{fig:FilterFunction}.

\bibliography{apssamp}

\end{document}